\documentclass[a4paper,12pt]{article}

\usepackage[utf8]{inputenc}
\usepackage[T1]{fontenc}
\usepackage{geometry}
\usepackage{amsmath}
\usepackage[normalem]{ulem}
\usepackage{amssymb,latexsym,amsmath}
\usepackage{setspace}
\usepackage{graphicx}
\usepackage{textcomp}
\usepackage[export]{adjustbox}
\usepackage{grffile}
\usepackage{gensymb}
\usepackage{float}
\usepackage{subfig}
\usepackage{natbib}
\usepackage{csvsimple}
\usepackage{hyperref}
\usepackage{verbatim}
\usepackage{multicol}
\usepackage{hyperref} 

\renewenvironment{thebibliography}[1]{%
\begin{oldthebibliography}{#1}%
\setlength{\parskip}{0ex}%
\setlength{\itemsep}{0ex}%
}%
{%
\end{oldthebibliography}%
}

\usepackage{mathptmx}
\usepackage{geometry}
\usepackage{xcolor} 

\begin{document}

\begin{center}
\large \textbf{
Simulations of linear polarization of precessing AGN jets at parsec scales
}
\end{center}

\begin{center}
R.V.~Todorov$^1$, E.V.~Kravchenko$^{1,2,4}$, I.N.~Pashchenko$^{2,5}$, and A.B. Pushkarev$^{3,2,6}$
\end{center}

\noindent$^1$Moscow Institute of Physics and Technology, Institutsky per. 9, Dolgoprudny, Moscow region, 141700, Russia\\
$^2$Astro Space Centre of Lebedev Physical Institute, Profsoyuznaya 84/32, Moscow 117997, Russia\\
$^3$Crimean Astrophysical Observatory, 298409 Nauchny, Crimea, Russia
\begin{center}
ORCIDs: $^4$ 0000-0003-4540-4095,  $^5$ 0000-0002-9404-7023,  $^6$ 0000-0002-9702-2307
\end{center}

\vspace{16pt}
\textit{Abstract.} The latest results of the most detailed analysis of multi-epoch polarization-sensitive observations of active galactic nuclei (AGN) jets at parsecs scales by very long baseline interferometry (VLBI) reveal several characteristic patterns of linear polarization distribution and its variability \citep{2023MNRAS.520.6053P, 2023MNRAS.523.3615Z}. Some of the observed profiles can be reproduced by a simple model of a jet threaded by a helical magnetic field.
However, none of the models presented to date can explain the observed polarization profiles with an increase in its degree towards the edges of the jet, and accompanied by a ``fountain'' type electric vector pattern and its high temporal variability in the center.
Based on simulations of the VLBI observations of relativistic jets, we show here that the observed transverse linear polarization profiles, atypical for the simple magnetic field models can be naturally produced assuming the finite resolution of VLBI arrays and precession of a jet on ten-years scales, observational indications of which are found in an increasing number of AGN.
In our simulations, we qualitatively reproduce the distribution of the electric vector and its variability, though the polarization images are characterized by a bright spine due to weak smearing, which is poorly consistent with observations. More effective depolarization can be obtained in models with the suppressed emission of the jet spine.\\

\vspace{6pt}

\vspace{4pt}
This is an author's translation of the original paper: Astronomicheskii Zhurnal, 2023, V. 100, N. 12, pp. 1132-1143, 
DOI: 10.31857/S0004629923120113, 
\href{https://sciencejournals.ru/view-article/?j=astrus&y=2023&v=100&n=12&a=AstRus2312011Todorov}{Journal link}

\section{Introduction}

There is a widely accepted view on a fundamental role of the magnetic field in the formation, acceleration and collimation of the relativistic jets in active galactic nuclei (AGN) \citep[AGN, see review][and references therein]{2019ARA&A..57..467B}. The degree of linearly polarized emission, produced by the synchrotron mechanism, can reach several tens of percent \citep{Lister_2005, galaxies5040093} in jet regions downstream from the radio core \footnote{By "radio core," we mean the visible base of the jet, typically the brightest and most compact feature on VLBI-maps.}, where the emission is optically thin \citep{1967ApJ...150..647P}. This, in addition to the the stable over a long time interval distribution of linear polarization indicates the presence of a global regular magnetic field in jets at parsec scales \citep{2017MNRAS.467...83K, 2023MNRAS.523.3615Z, 2023MNRAS.520.6053P}.

Polarimetric Very Long Base Interferometry (VLBI) is used as the most effective diagnostics of the magnetic field and the jet structure due to the high angular resolution \citep[see for example][]{2020AdSpR..65..712B}.
In the polarimetric observations of the AGN samples, the predominantly bimodal distribution of the Electric Vector Position Angle (EVPA, ${\rm tg} (2\chi) = U/Q$, where $Q$ and $U$ are the Stokes parameters) was found: perpendicular and parallel to the jet direction \citep{2000MNRAS.319.1109G, 2023MNRAS.520.6053P}.
This picture can be explained by the model of helical magnetic field \citep{2005MNRAS.360..869L}, which is formed by the differential rotation of the accretion disk or ergosphere of a black hole.

There is an substantial evidence of a presence of the helical magnetic field detected on sub-parsec \citep{2016ApJ...817...96G, 2020ApJ...893...68K, 2021A&A...648A..82P}, parsec \citep{2009MNRAS.393..429O,2012AJ....144..105H, 2017MNRAS.467...83K}, and kiloparsec scales in AGN jets \citep{2015A&A...583A..96G, 2021ApJ...923L...5P}.
In addition, some observational studies demonstrate the transverse gradient of the Faraday rotation measure \citep{2004ApJ...612..749Z, algaba13}, which serves as a most convincing evidence for the presence of a toroidal component of the magnetic field \citep{2002PASJ...54L..39A, zamaninasab_etal13, 2015MNRAS.450.2441G, 2021ApJ...910...35L}.
Furthermore, depending on the jet viewing angle and the helical pitch angle, the helical magnetic causes a "spine-sheath" polarization structure, with longitudinal polarization angles along the central ridgeline of the jet (spine), surrounded by regions with transverse EVPAs closer to jet edges (sheath) \citep{1999ApJ...518L..87A, 2005MNRAS.356..859P, 2005MNRAS.360..869L}.

\subsection{Observed transverse polarisation profiles}

Studies indicate that AGN jets can change its direction over time \citep{2006ApJ...647..172S, 2013AJ....146..120L}.
This can result in the true transverse profile of the jet being revealed only after accumulating (exposure increase, stacking) observations carried out over many years \citep{2017MNRAS.468.4992P, 2020MNRAS.tmp.1273K}. Analysis shows that typically observations over an about 6 years time interval are sufficient to fill the true jet cross-section.
For example, in the quasar 3C~273, changes in the jet direction resulted in variations in the observed parsec-scale transverse Faraday rotation gradient \citep{2021ApJ...910...35L}. This is consistent with a scenario in which previously invisible parts of the external wide Faraday screen are illuminated.

Recently, the MOJAVE group \footnote{Monitoring Of Jets in Active galactic nuclei with VLBA Experiments \citep{2018ApJS..234...12L}, https://www.cv.nrao.edu/} presented the largest and most detailed study of the polarization properties of AGN jets to date \citep{2023MNRAS.520.6053P}.
This study is based on multi-epoch long-term monitoring using the Very Long Baseline Array (VLBA) at a frequency of 15\,GHz, during which averaged multi-epoch (stacked) maps of the linear polarization distribution of 436 radio-loud AGNs were obtained.
The number of individual epochs of observations for each source ranged from 5 to 139, with a median of 9.
The temporal coverage for each individual source ranged from one year to 24 years, with a median of 7 years \citep{2023MNRAS.523.3615Z}.
\citet{2023MNRAS.520.6053P} analyzed the distribution of the fractional polarization and polarization angle on the stacked maps. The accompanying study \citep{2023MNRAS.523.3615Z} also investigated the variability of the linear polarization on timescales of about a decade.
Stacking allowed for a several-fold increase in the image sensitivity for sources observed in the typical number of epochs and by an order of magnitude for the most frequently observed sources, reaching values of $\sim 30~\mu$Jy~beam$^{-1}$, enabling to reveal more complete total and polarized brightness distributions. \citet{2023MNRAS.520.6053P} and \citet{2023MNRAS.523.3615Z} show that the stable and most complete linear polarization distribution over the source is achieved on time interval of 10 years of VLBI observations.

\begin{figure}
\includegraphics[width=0.49\columnwidth]{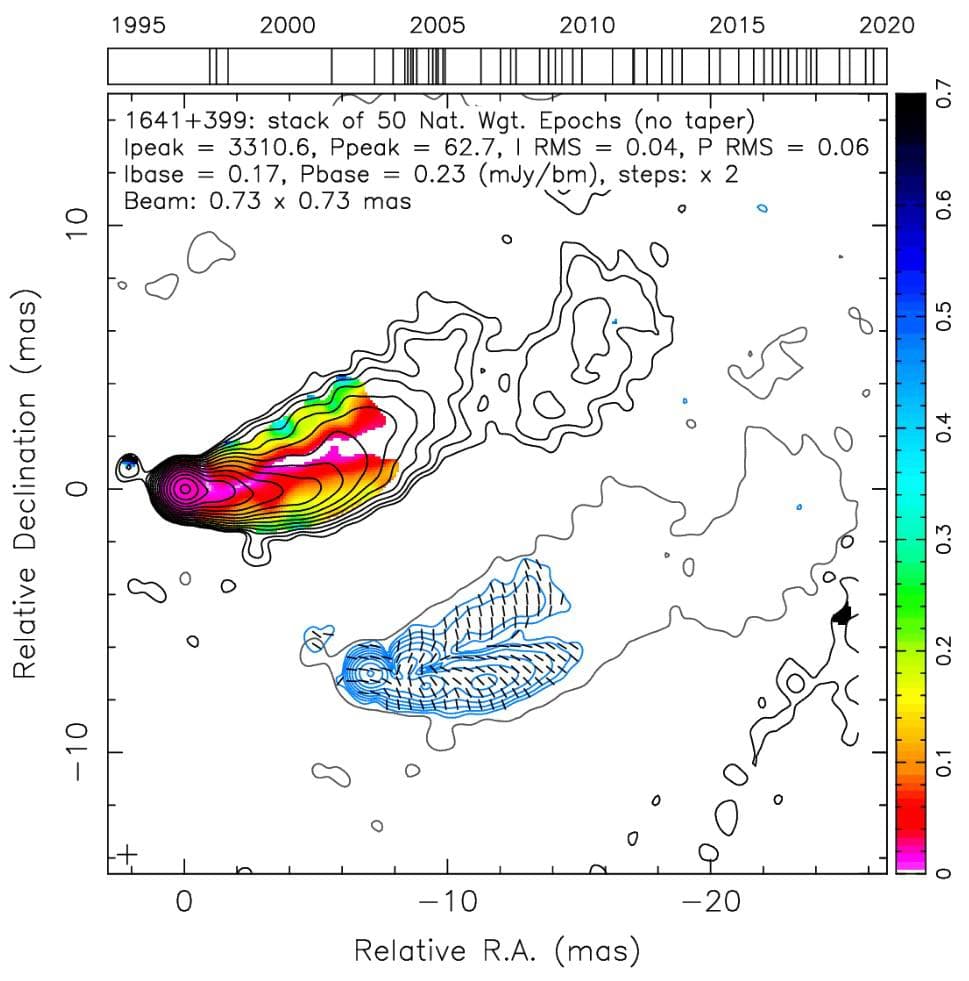}
\includegraphics[width=0.49\columnwidth]{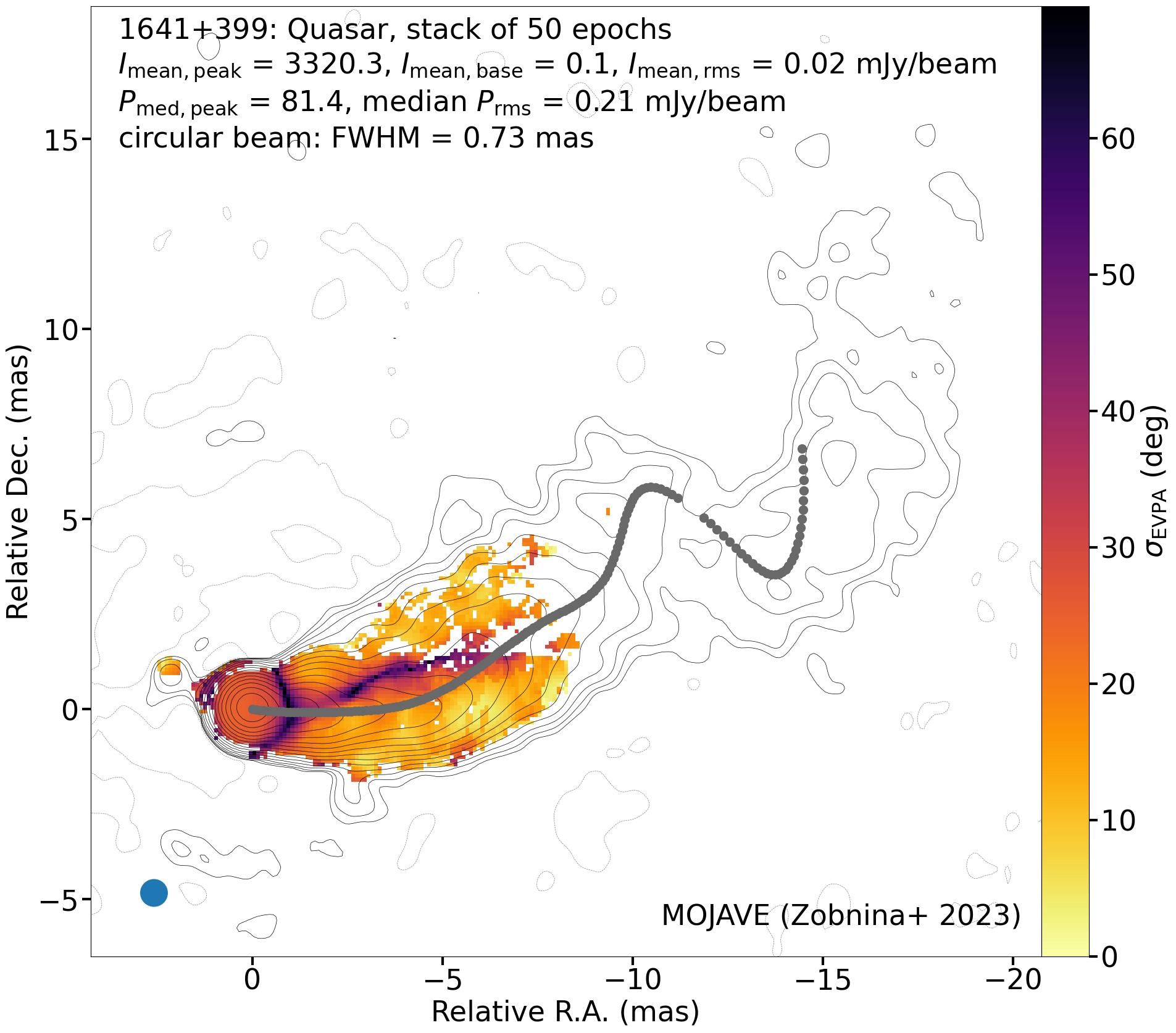}
\caption{(Left) The 15\,GHz stacked linear polarization image of the quasar 3C~345 \citep{2023MNRAS.520.6053P}. The fractional polarization is shown in color, the total intensity is represented by black contours, the linear polarization is depicted by blue contours, and the EVPA is indicated by dashes, without correction for the Faraday rotation.
(Right) Image of the EVPA variability (color) on top of the total intensity contours \citep{2023MNRAS.523.3615Z}. The gray dots denote the stacked total intensity ridgeline.
\label{fig:3c345obs}}
\end{figure}

This stacking analysis shows that for most sources, the fractional polarization orthogonal to the jet direction increases towards its edges, thus having a U-shaped profile \citep{2023MNRAS.520.6053P}. In rare cases, the fractional polarization has a peak in the jet center forming a W-shaped transverse profile.
\citet{2023MNRAS.520.6053P} identified three typical transverse profiles of the EVPAs: their orientation along the ridge line, transverse, and a fountain-like distribution. The latter case occurs on average in every fifth source with non-zero polarization in the jet region downstream the core (e.g., 0234$+$285, 0336$-$019, 0945$+$408, etc.). This pattern is characterized by an EVPA parallel to the local direction along the jet spine with a gradual rotation of $90^{\circ}$ toward the jet edges (Fig.~\ref{fig:3c345obs}).
In this case, the transverse EVPA at the jet edges is accompanied by an increase in the degree of polarization, while the longitudinal EVPA at the jet center exhibits a lower degree of polarization.
Some of these sources show significant temporal variability in the EVPA direction along the jet spine \citep{2023MNRAS.523.3615Z} with an amplitude of about $40^{\circ}-50^{\circ}$ (e.g., 0336$-$019, 0735$+$178, 1641$+$399, 1920$-$211, and 2251$+$158) (see Fig.~\ref{fig:3c345obs}). At the same time, the median standard deviation $\sigma_{\rm EVPA}$ for extended jet regions in the entire sample of 436 AGN is substantially lower, about 10 degrees.
Faraday rotation at 15\,GHz for the AGN jets in the MOJAVE sample typically reaches only a few degrees and is mostly observed in the core regions \citep{2012AJ....144..105H, 2017MNRAS.467...83K}. Therefore, its influence on the polarization properties observed at 15\,GHz can be neglected.

\subsection{Existing models of linear polarization of AGN jets}

The first analytical calculations of transverse polarization profiles for the AGN jets threaded by a large-scale helical magnetic field were presented by \citet{2005MNRAS.360..869L}.
\citeauthor{2005MNRAS.360..869L} showed that the helical magnetic field can explain almost all the polarization profiles observed at that time. For example, with a small pitch angle, longitudinal component of the magnetic field will prevail, and in the case of a strongly twisted magnetic field, the orientation will be predominantly orthogonal with respect to the jet direction. This will correspondingly affect the direction of polarization of the observed synchrotron emission.
The more complex polarization configuration ``spine-sheath'' is also explained by the helical field \citep{2014MNRAS.444..172G}.
Depending on the pitch and the jet viewing angles, asymmetric profiles can also appear \citep{2011MNRAS.415.2081C}, in which longitudinal and orthogonal components of the magnetic field prevail on both or one side of the jet \citep{2021Galax...9...58G}.

In more complex analytical models, W-, U-shaped profiles of the degree of polarization and EVPA can also be obtained using a toroidal or strongly twisted large-scale magnetic field \citep{2008ApJ...679..990Z}.
\citet{2013MNRAS.430.1504M} considered the least model-dependent case of a purely helical field and a homogeneous jet, focusing on the main characteristics of the transverse profiles of total and polarized intensities. Assuming different pitch and jet viewing angles, \citet{2013MNRAS.430.1504M} successfully reproduced asymmetric polarization profiles.

Recently, a more complex model has been proposed to explain the observed kinematics of the individual components and polarization profiles of the AGN jets \citep{2018ARep...62..116B, 2023MNRAS.520.6335B}.
Butuzova et al. consider a jet model in which its axis forms a spiral on the surface of a cone, and the assumed configuration of the magnetic field is either helical, with different pitch angles, or toroidal in the central part of the jet and poloidal in its sheath.
\citet{2023MNRAS.520.6335B} obtained a qualitative correspondence between the model and observed transverse profiles simultaneously for total and polarized intensities, U-shaped polarization profiles and the deviation of EVPA from the local jet axis for both the helical magnetic field and the ``spine-sheath'' configuration \citep{2022evlb.confE...5B, 2023MNRAS.520.6335B}.

At the same time, none of the models presented to date can explain the observed U-shaped polarization profile, accompanied by a ``fountain''-like EVPA pattern and its high variability along the jet axis.
In this paper, we present a model of a precessing jet with a strongly twisted helical or toroidal magnetic field and produce simulated polarization-sensitive VLBI images of the AGN jets for their direct comparison with observations.

\section{Model of a precessing jet and simulations of multi-epoch VLBI observations}
\label{sec:model_and_simulations}

We are considering a model of a conical jet with a toroidal or highly twisted helical magnetic field with a typical transverse polarization structure characterized by three peaks: a longitudinal polarization at the center and weaker transverse polarizations at the jet edges \citep{1981ApJ...248...87L, 2013MNRAS.430.1504M}. In the case of homogeneous heating (acceleration of emitting particles) of the plasma across the jet, the amplitude of the central peak is always two to three times higher than that of the side peaks. Such cases occur for pitch angles of the helical magnetic field $\psi' \geq 70^{\circ}$ and small inclination angles of the jet axis to the line of sight $\theta' \leq 30^{\circ}$, where all prime angles are in the plasma frame. Thus, the angle in the plasma frame $\theta' = 90^{\circ}$ corresponds to the angle in the observer's frame $\theta_{\rm LOS} = 1/\Gamma$. The results of Monte Carlo simulations on a flux-based statistically complete subsample of quasars 1.5JyQC MOJAVE, using the model of a constant-speed jet \citep[Fig. 11 in][]{2019ApJ...874...43L}, predict a typical value of the Lorentz factor of the jet $\Gamma = 10$ \citep{2019ApJ...874...43L}, which corresponds to $\theta_{\rm LOS} \thicksim 5^{\circ}$, with the most probable value of $\theta_{\rm LOS} \thicksim 2^{\circ}$. At such small viewing angles, even slight "swinging" of the jet due to, for example, precession, is enhanced by projection effects.

Assuming the orientation of the jet changes over time in a way that the jet axis forms a conical surface with the vertex at the central engine of the source. In this case, the jet viewing abgle is always greater than the sum of its half-opening angle and the precession angle, i.e. the observer never looks inside the outflow.
Due to the small jet viewing angle, even slight real changes in the jet direction will result in significant changes in its positional angle (PA) in the projection on the image plane.
The amplitude of the jet oscillations ($\eta$) at the level of $\pm0.2^{\circ}$ corresponds to a visible amplitude of $\pm15^{\circ}$ as a result of the projection. Due to the change of the jet position angle in the sky projection, and then averaging all corresponding images, the observed polarized intensity at the jet center will be smeared due to the superposition of polarization with different orientations. This naturally causes increased variability of the EVPA in this region.
Significant variations in the position angle of jet components observed in the considered sources \citep{2021ApJ...923...30L}, their large apparent jet opening angles \citep{2017MNRAS.468.4992P}, and the resolved transverse polarization structure serve as the basis for the proposed scenario.

For the simulation of VLBI-observations, we consider a model of a relativistic jet with a Lorentz factor $\Gamma$ \citep{1979ApJ...232...34B, 1981ApJ...243..700K}. The velocity field is central, and the velocity of the plasma bulk motion is directed away from the central engine. This is important for small inclination angles of the jet axis to the line of sight. The jet has a half-opening angle $\phi$ and is observed at an angle $\theta$. The redshift is chosen such that the integrated flux on the model image is of the order of 1\,Jy, which corresponds to the typical flux of real observations.
We also assume an isotropic power-law distribution of emitting particles $N(\gamma) \propto \gamma^{-s}$ with $\gamma_{\rm min} = 10$ and power $s = 2.5$ \citep[which corresponds to a spectral index\footnote{The spectral index is determined by the dependence of the flux density $S$ on the frequency $\nu$ as $S\propto\nu^{\alpha}$.} $\alpha = -0.75$][]{2014AJ....147..143H}, in an equipartition distribution with helical magnetic field. We use $\gamma_{\rm max} = 10^4$, although the exact value of this parameter is not essential.
We used an electron-proton plasma, which leads to a nonzero Faraday rotation effect, with the rotation measure reaching values of 200 $\text{rad}/\text{m}^2$. This magnitude leads to a rotation of EVPA of less than $10^\circ$ at 15\,GHz.

The solution to the transfer equations of polarized synchrotron radiation is calculated taking into account self-absorption and Faraday effects for Stokes $I$, $Q$, $U$, $V$ as follows \citep[for example,][]{2002A&A...388.1106B}:

\begin{equation}
\begin{gathered}
\frac{d I}{d l}=\eta_I-\kappa_I I-\kappa_Q Q-\kappa_U U-\kappa_V V \\
\frac{d Q}{d l}=\eta_Q-\kappa_I Q-\kappa_Q I-\kappa_F U-h_Q V \\
\frac{d U}{d l}=\eta_U-\kappa_I U-\kappa_U I+\kappa_F Q-\kappa_C V \\
\frac{d V}{d l}=\eta_V-\kappa_I V-\kappa_V I+h_Q Q+\kappa_C U
\end{gathered}
\end{equation}
where $\eta_I , \eta_U , \eta_Q ,\eta_V$ are the emissivity coefficients, $\kappa_I , \kappa_U ,\kappa_Q , \kappa_V$ are the absorption coefficients, $\kappa_F$ is the Faraday rotation coefficient, and $\kappa_C$ and $h_Q$ are rotation coefficients for Faraday conversion.
The unit direction vector along the line of sight $\hat{n}_{\rm LOS}$ is transformed into the moving plasma frame (relativistic aberration), as
\begin{equation}
\hat{n}_{\rm LOS}' = D \hat{n}_{\rm LOS} - (D + 1) \frac{\Gamma}{\Gamma+1} \beta,
\end{equation}
where the Doppler factor $D = (\Gamma (1 - \upsilon \cdot \hat{n}_{\rm LOS} /c))^{-1}$ \citep[See for example][]{2009ApJ...695..503G}.
Connection of the angles in a co-moving and observer's frame are given by the relation:
\begin{equation}
\Gamma[1 - \beta \cos \theta]\sin \theta' = \sin \theta.
\label{eq:ang_abberation}
\end{equation}

For direct comparison of a model linear polarization images with an analogous observed maps, we create synthetic single-epoch maps that correspond to different phases of the jet rotation.
For this, we first transform the model brightness distribution into the spatial frequency plane $(u, v)$ using Fourier transform, with the $(u, v)$-coverage taken from the real observations \footnote{We used 36 epochs of observations of the source 3C~345 (1641$+$399).}. In other words, we generate a set of synthetic interferometric visibilities. Then, thermal noise, calculated based on the scatter of the actual observation visibilities \citep{briggs}, is added.
Next, we use the CLEAN procedure \citep{CLEANref,1980A&A....89..377C} implemented in the package \texttt{Difmap} \citep{difmap} to produce linear polarization images from the synthetic visibilities for each of the stokes $I$, $Q$, and $U$.
For a convolution, a circular 0.73\,mas beam size is used, which corresponds to one used to construct the stacked map of the source 1641+399 \citep{2023MNRAS.520.6053P}.
Then, the set of synthetic polarization maps is averaged to create stacked maps in the same way as was done for the real multi-epoch observation data \citep{2023MNRAS.520.6053P}.
Specifically, stacked maps are first formed for the stokes $U$ and $Q$, and then for the polarization $P = \sqrt{U^2 + Q^2}$ and the polarization angle $\theta = \frac{1}{2} \arctan\left(\frac{U}{Q}\right)$ are created.

\section{Results}

\begin{figure}
\includegraphics[width=0.5\columnwidth]{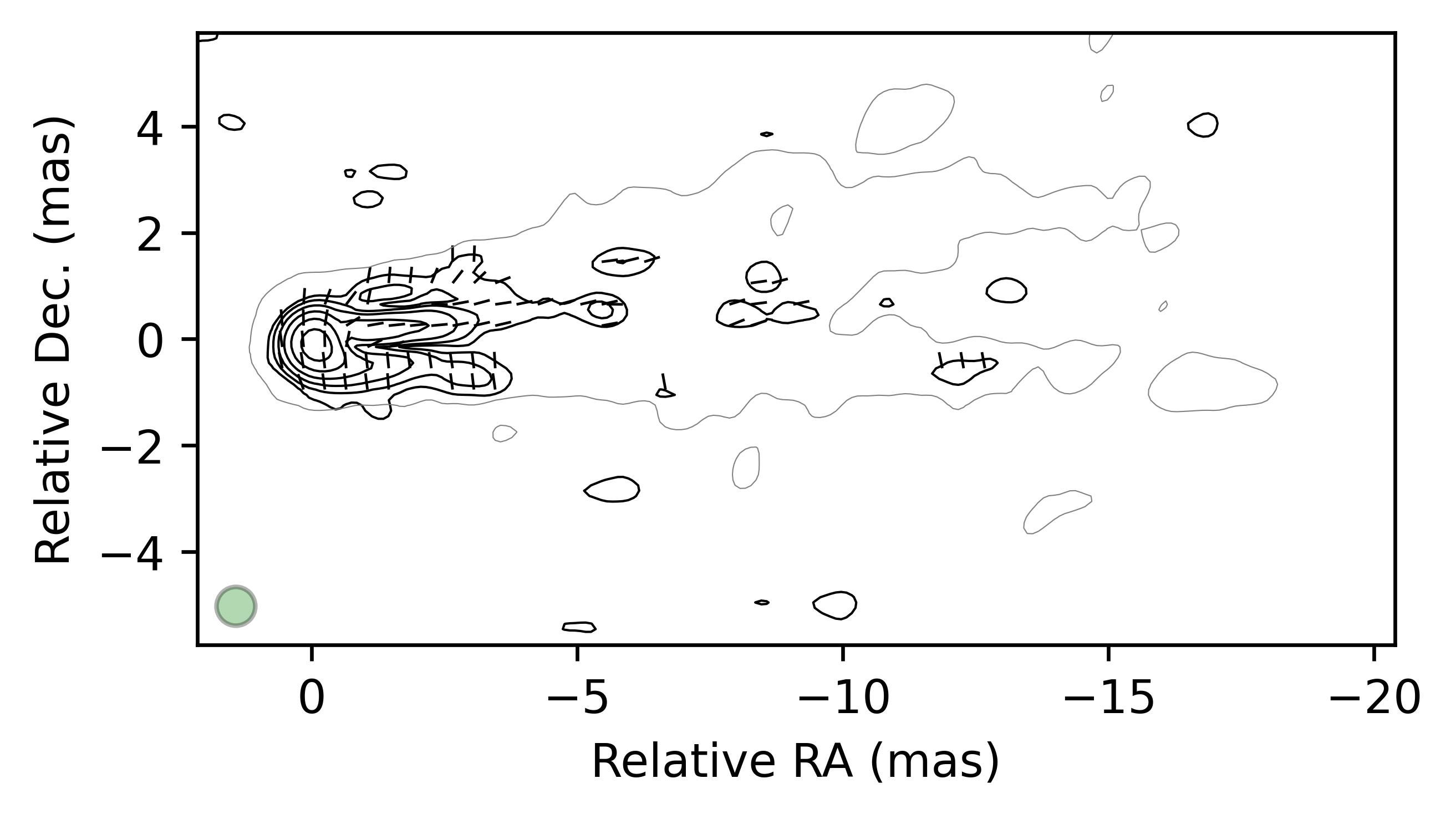}
\includegraphics[width=0.5\columnwidth]{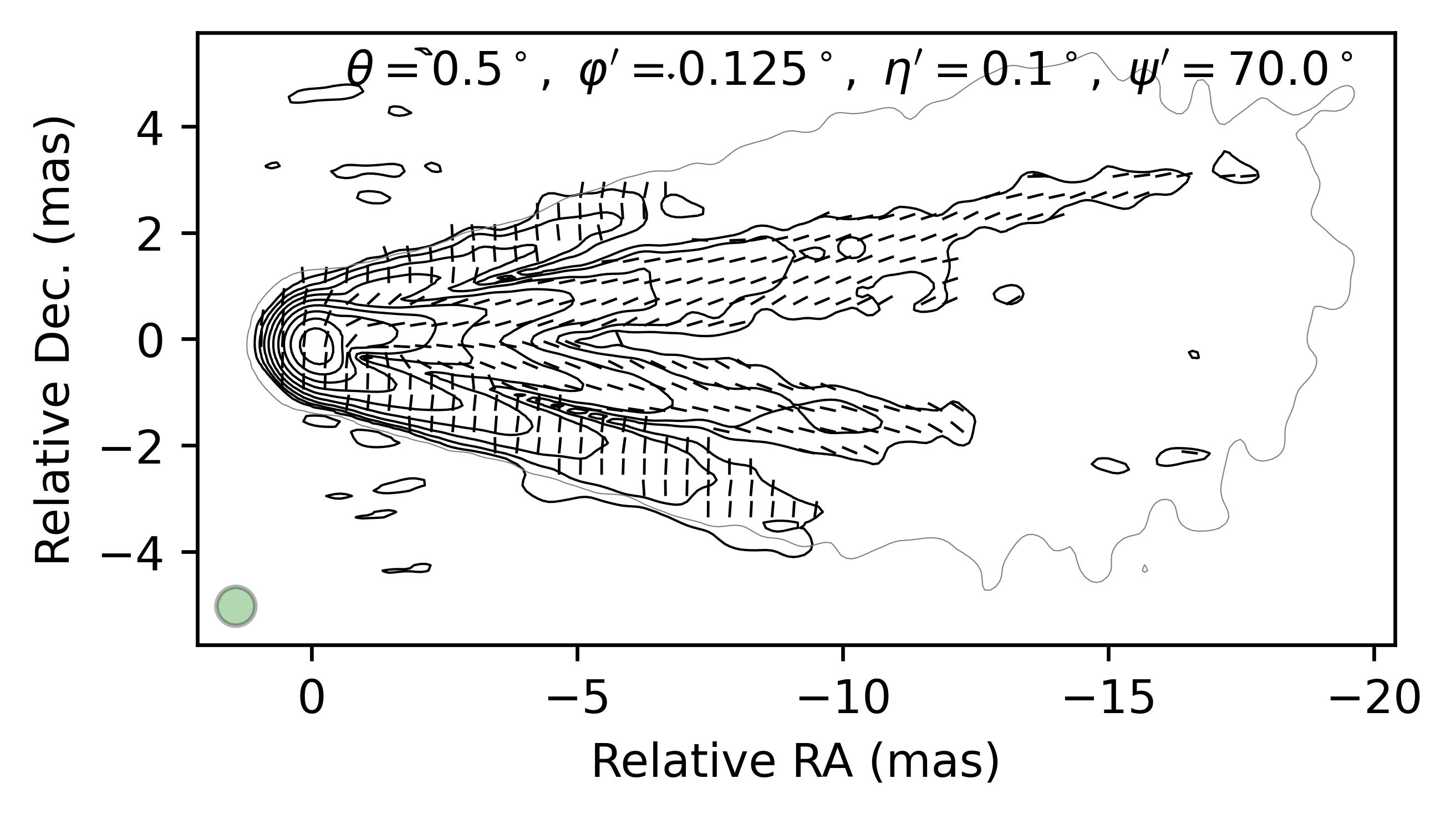}
\includegraphics[width=0.5\columnwidth]{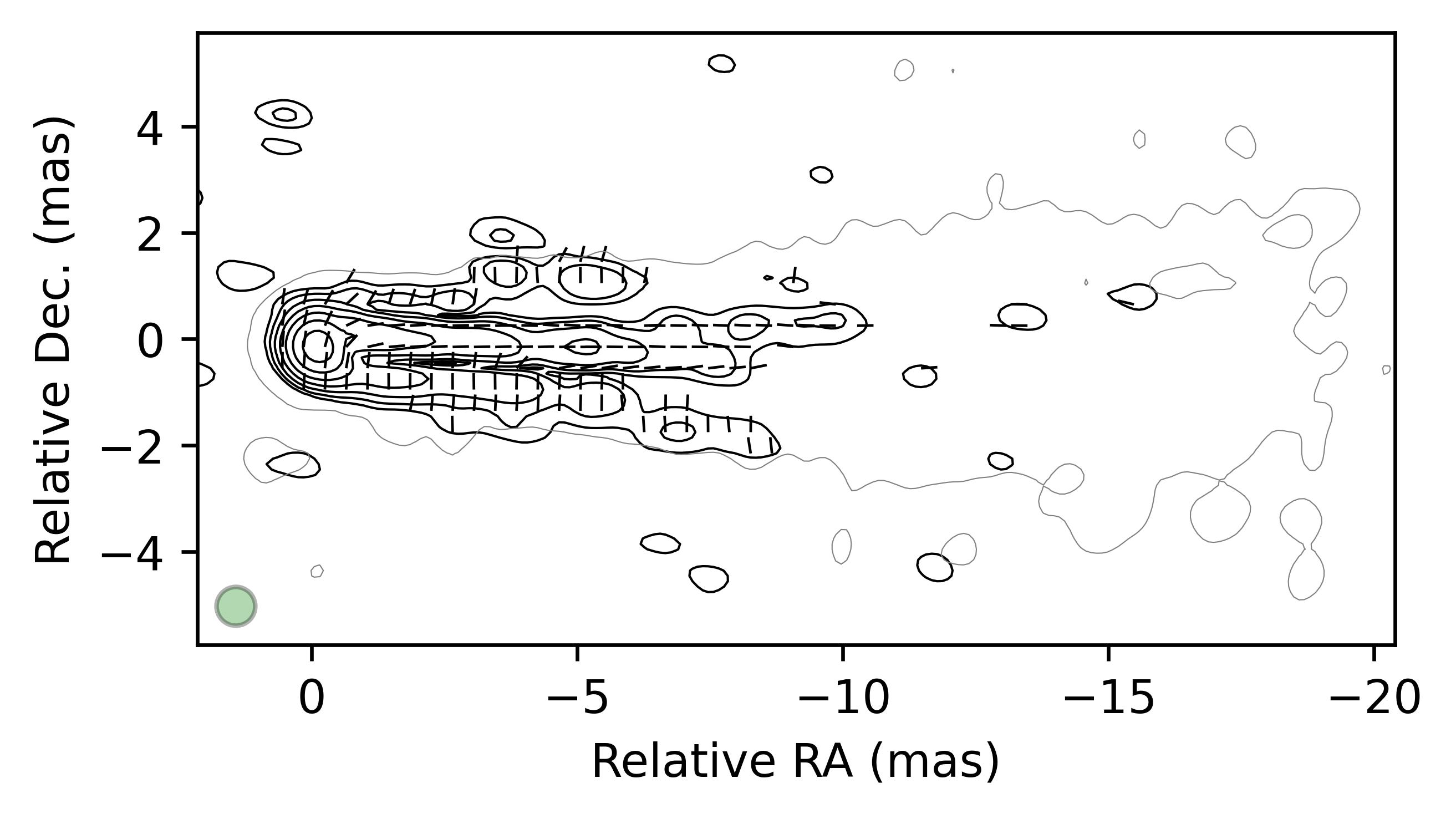}
\includegraphics[width=0.5\columnwidth]{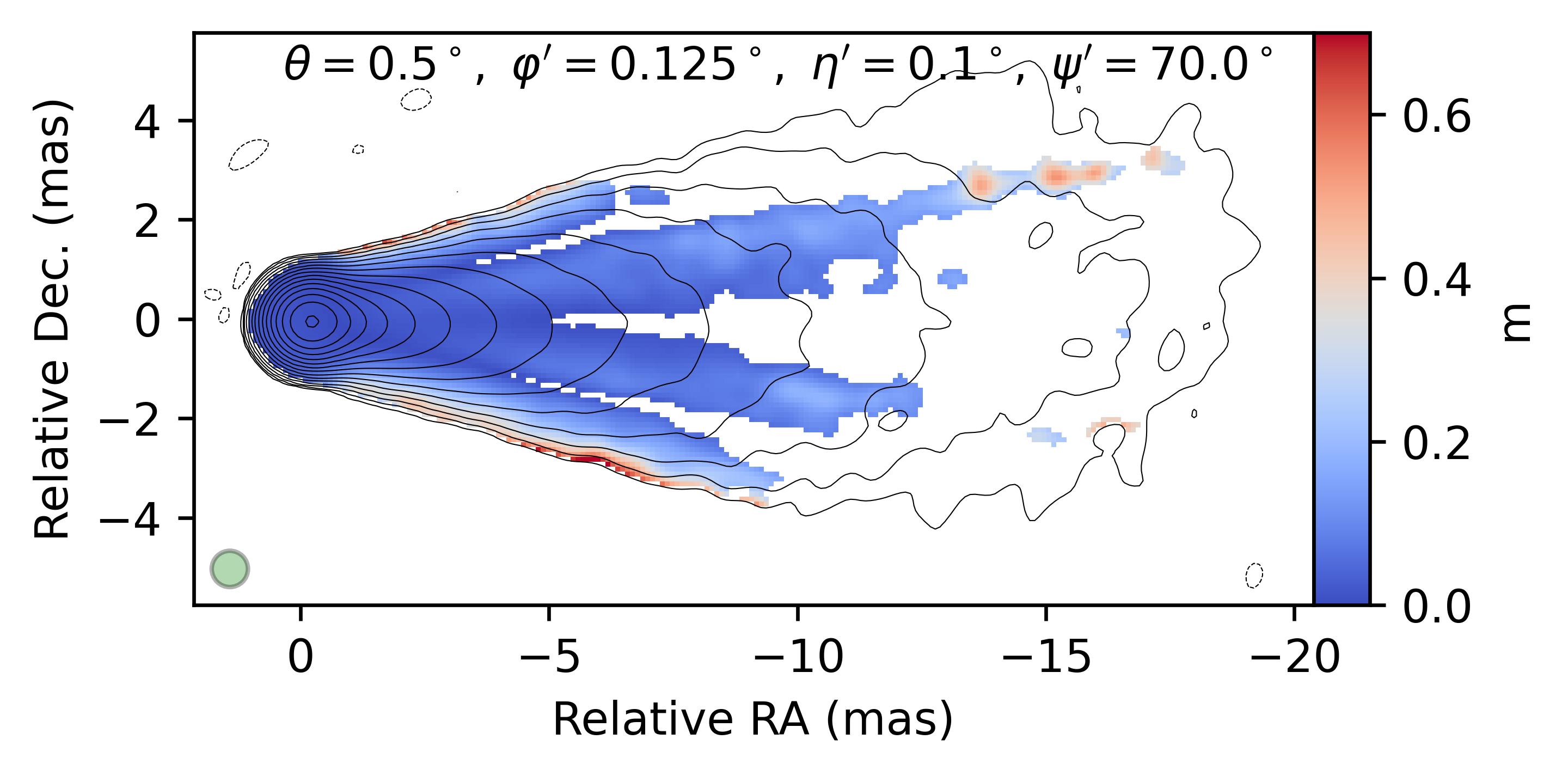}
\includegraphics[width=0.5\columnwidth]{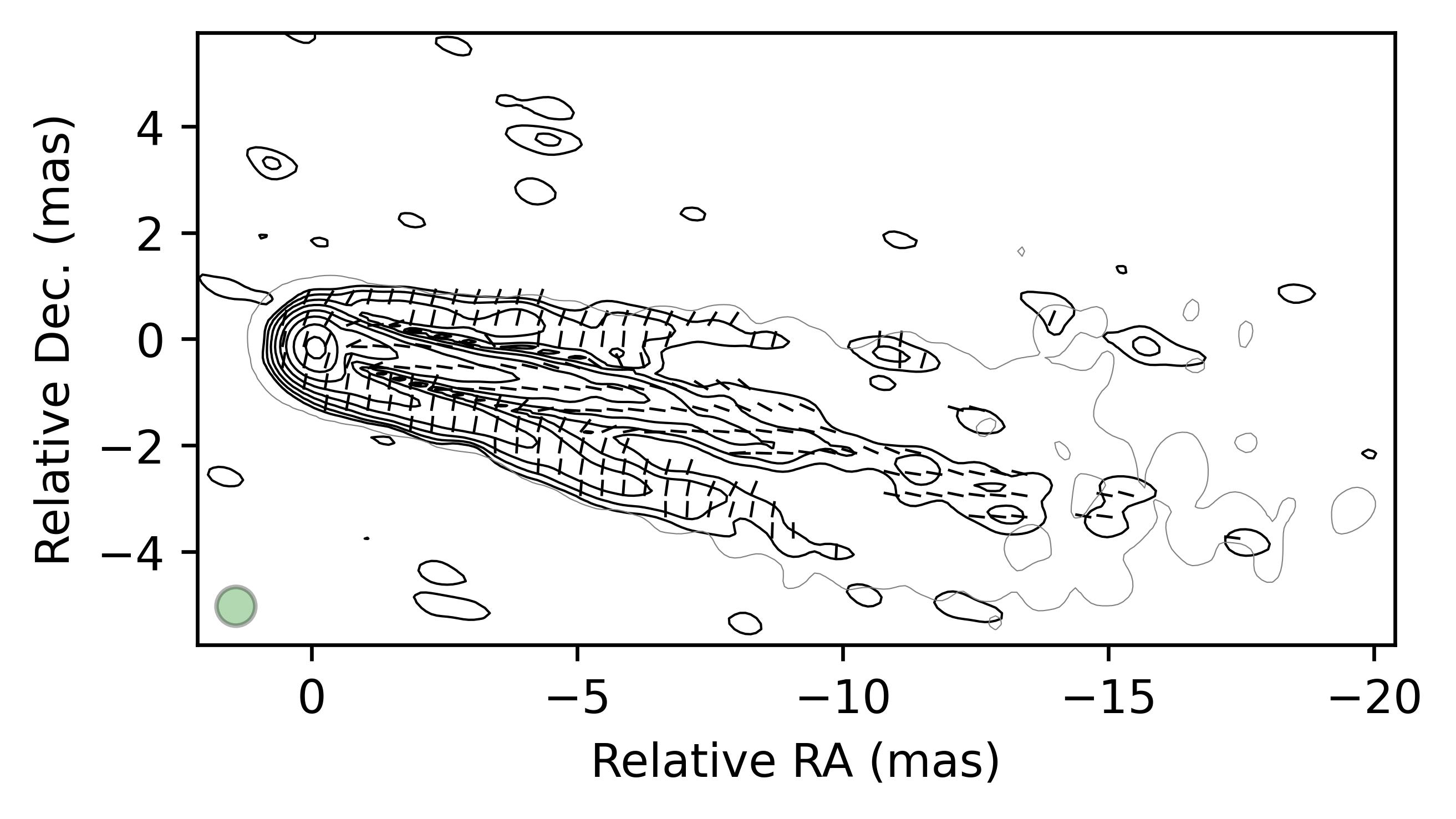}
\includegraphics[width=0.5\columnwidth]{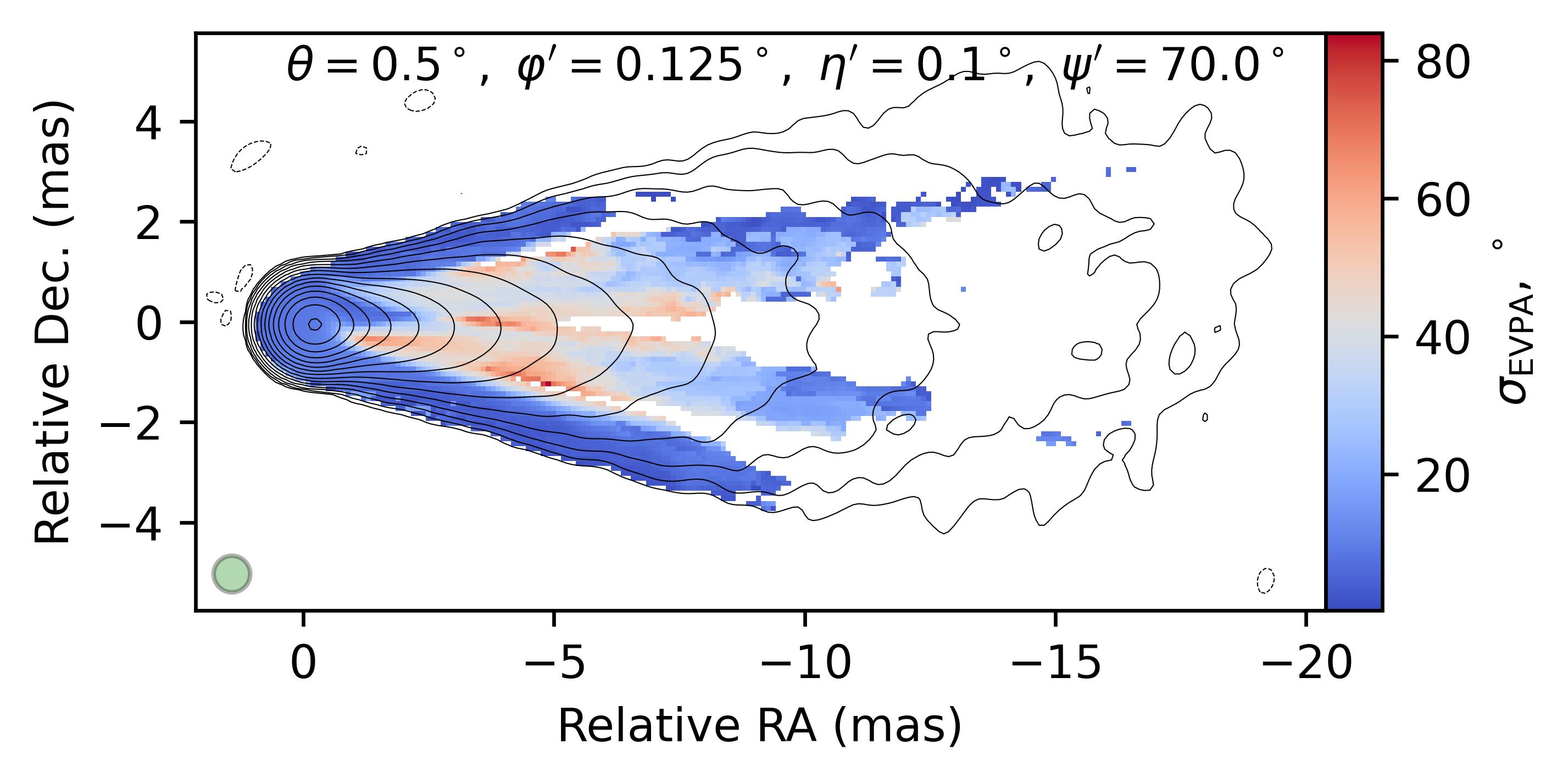}
\caption{Synthetic single-epoch (left) and stacked (right) linear polarization jet maps.
The single-epoch images correspond to different phases of the jet rotation. The following model parameters were used: a conical jet with a true half-opening angle of $\phi'=0.125^\circ$, with its precession axis inclined to the line of sight by $\theta=0.5^\circ$, precessing with an amplitude of $\eta'=0.1^\circ$; the pitch angle of the helical magnetic field is $\psi'=70^\circ$.
Top-right -- map of the EVPA and polarised intensity distribution, middle-right -- map of the degree of linear polarization, bottom-right -- map of the variability of the EVPA ($\sigma_{\rm EVPA}$). For the convolution, circular beam was used with a FWHM beam size of 0.73\,mas, shown by the green circle in the bottom-left corner. It corresponds to the beam used to generate the stacked map of the source 1641$+$399 \citep{2023MNRAS.520.6053P}. The maps were obtained by averaging 36 single-epoch images.
The total flux density is 0.7\,Jy, and the value of the linearly polarized flux density on the stacked map is approximately 50\,mJy.
Gray contours of the total intensity are drawn at $3\sigma_{\rm I}=10$ mJy/beam and increase by a factor of 2. Black contours of the linearly polarized intensity are given at $3\sigma_{\rm P}=0.18$ mJy/beam and also increase by a factor of 2.
\label{fig:sim_vlbi_jet}}
\end{figure}

In this study, we considered pitch-angles of $70^\circ$, $80^\circ$, and $89^\circ$, i.e., predominantly toroidal magnetic field.
To create a synthetic stacked image, 36 positions corresponding to real epochs of observations of the source  1641$+$399\footnote{https://www.cv.nrao.edu/mojave/sourcepages/1641+399.shtml} and various initial phase rotations of the jet axis along the circular cone with an amplitude ranging from 0.1$^\circ$ to 0.15$^\circ$ were selected. Each position corresponds to one observational epoch, and examples of the obtained VLBI-maps are shown in Figure \ref{fig:sim_vlbi_jet}. 
The half-opening angle of the conical jet was chosen as $\theta/4$, which corresponds to an apparent half-opening angle of $\sim 10^\circ$ \citep{2017MNRAS.468.4992P}. 
The inclination angles of the jet axis to the line of sight $\theta =$ 0.3$^\circ$, 0.5$^\circ$, 1$^\circ$, and 2$^\circ$ from the observer's point of view were considered, which correspond to the angles $\theta' =$ 6$^\circ$, 10$^\circ$, 20$^\circ$ and 38$^\circ$ in the plasma's frame, respectively (See Eq.~\ref{eq:ang_abberation}).

\begin{figure}
\includegraphics[width=0.47\columnwidth]{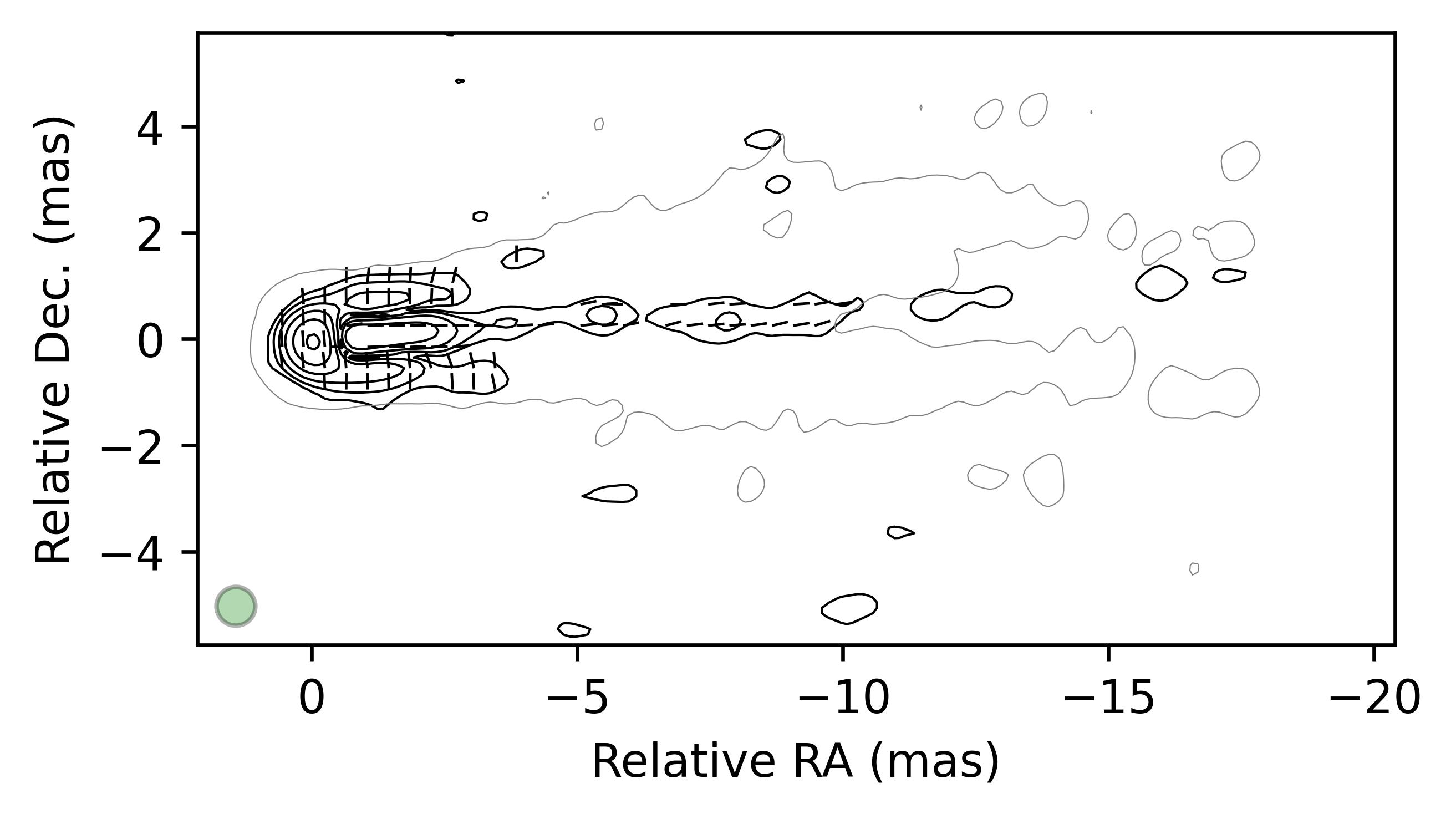}
\includegraphics[width=0.49\columnwidth]{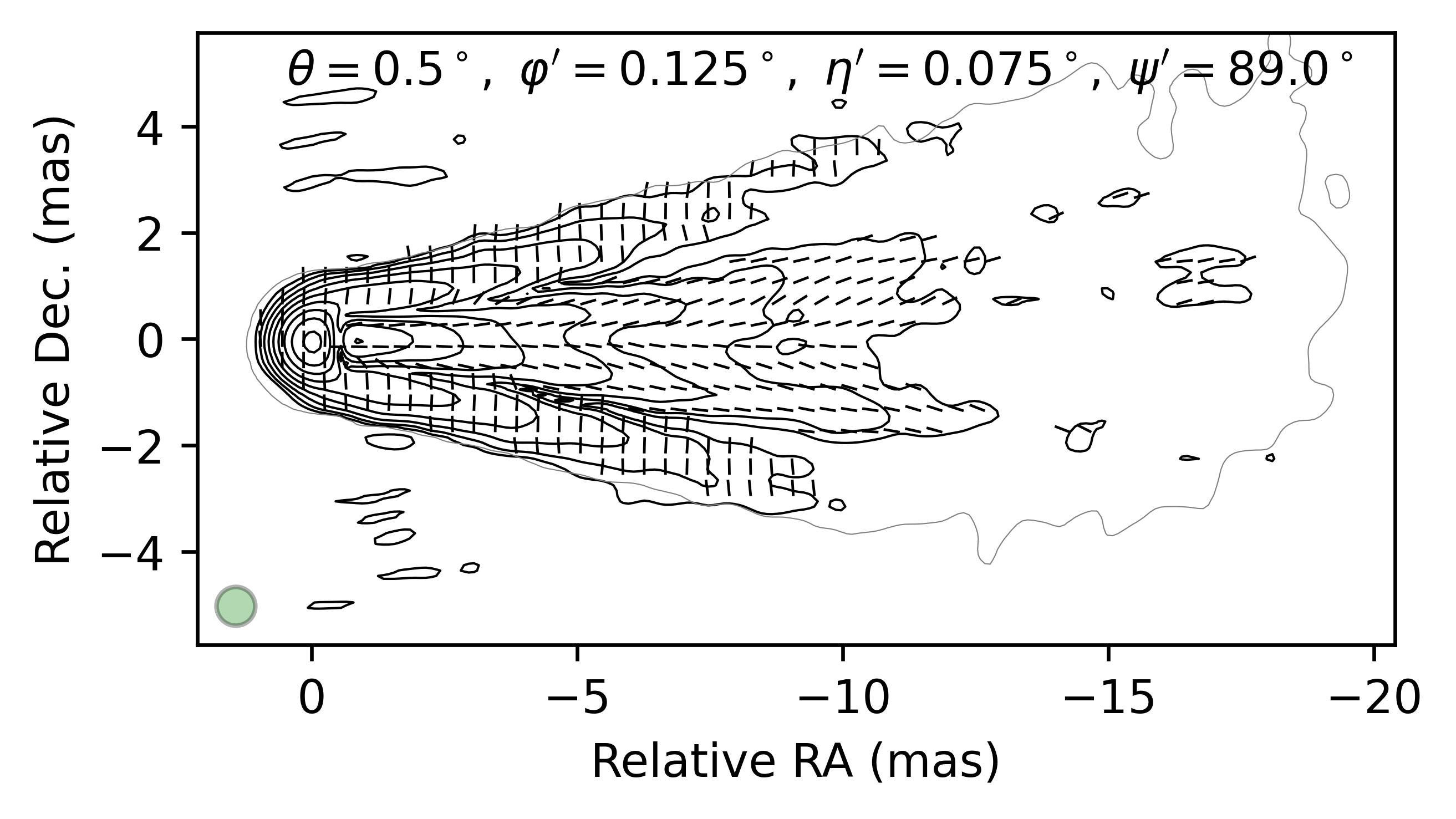}
\includegraphics[width=0.47\columnwidth]{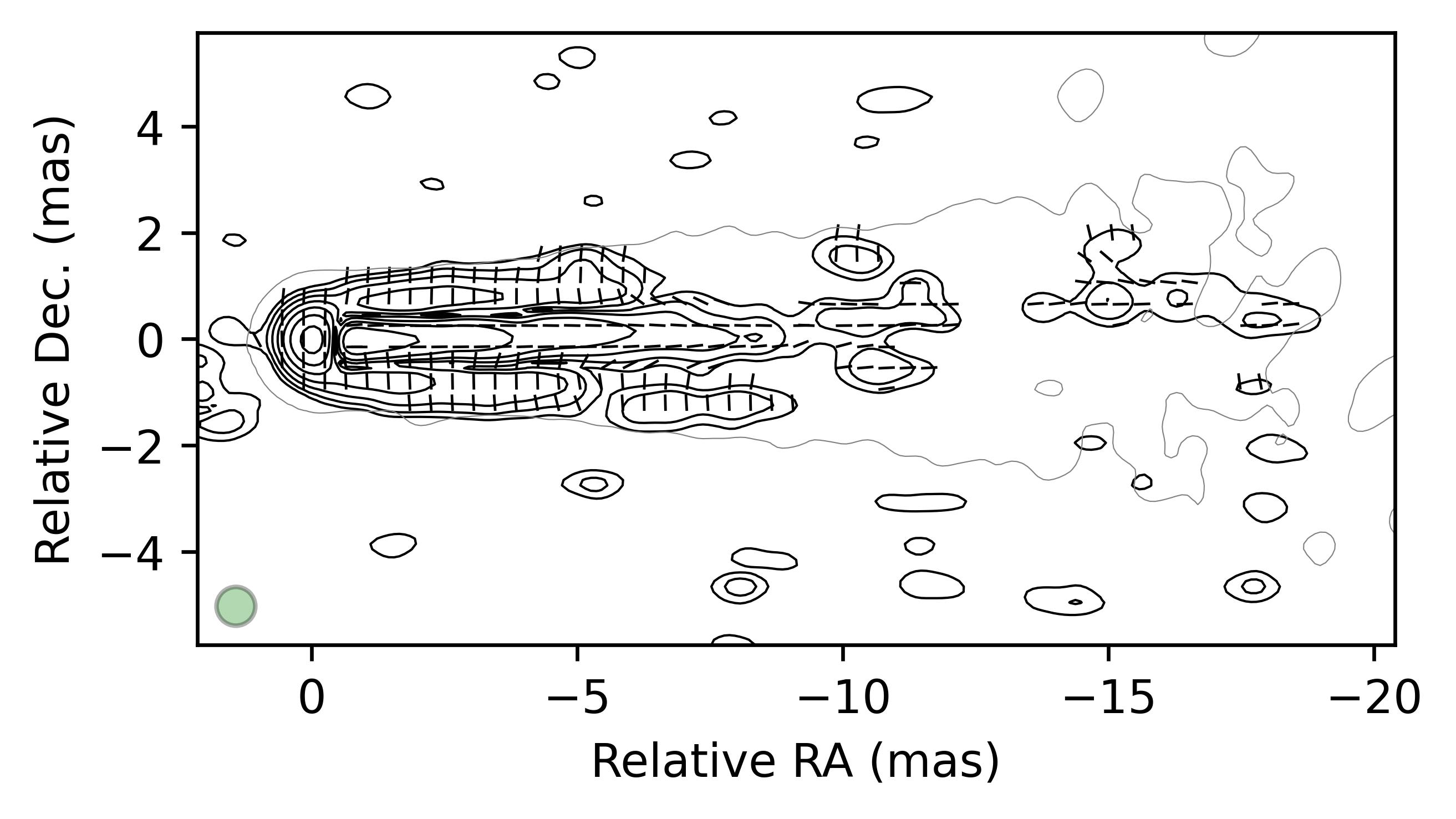}
\includegraphics[width=0.50\columnwidth]{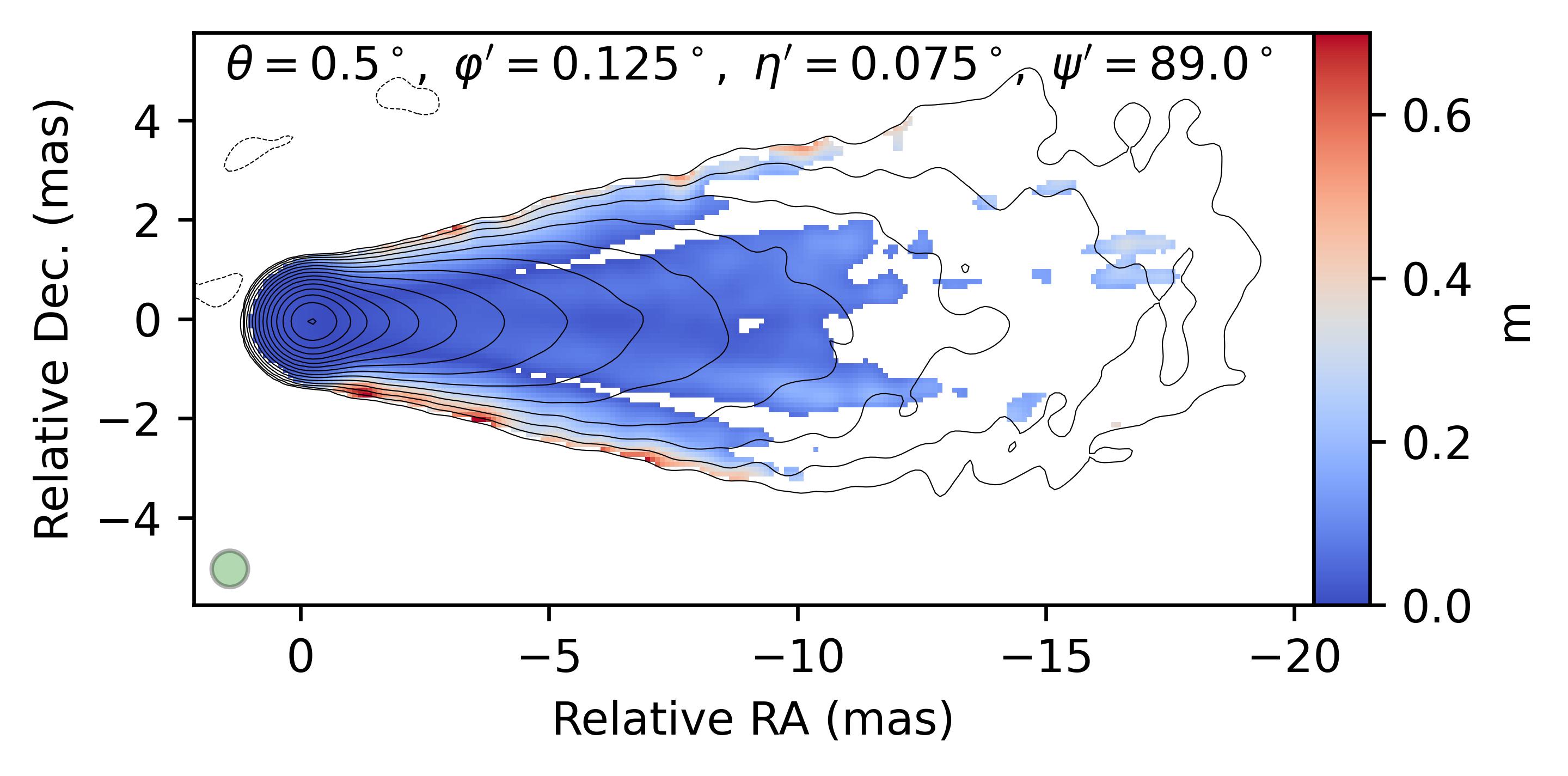}
\includegraphics[width=0.47\columnwidth]{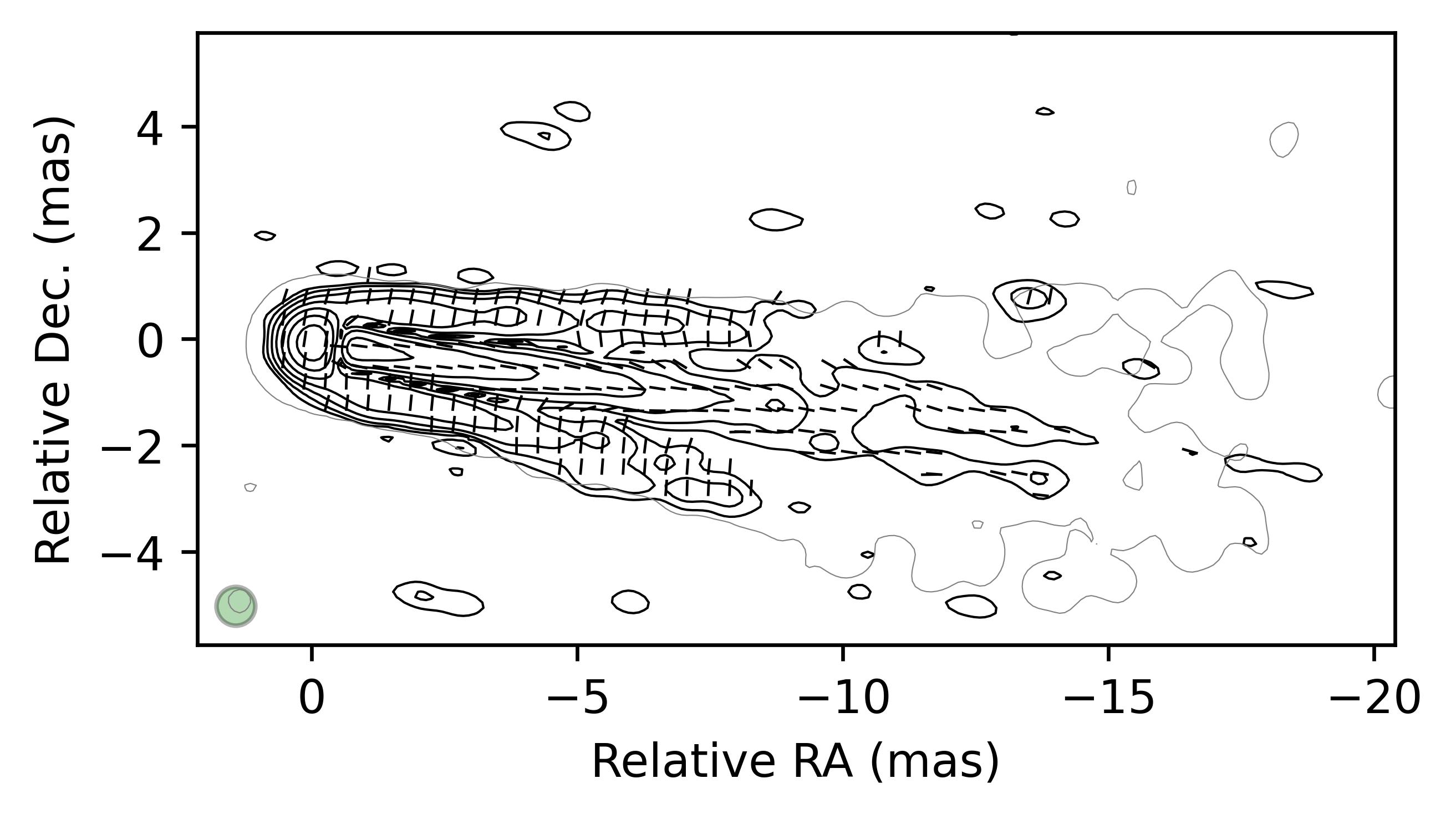}
\includegraphics[width=0.51\columnwidth]{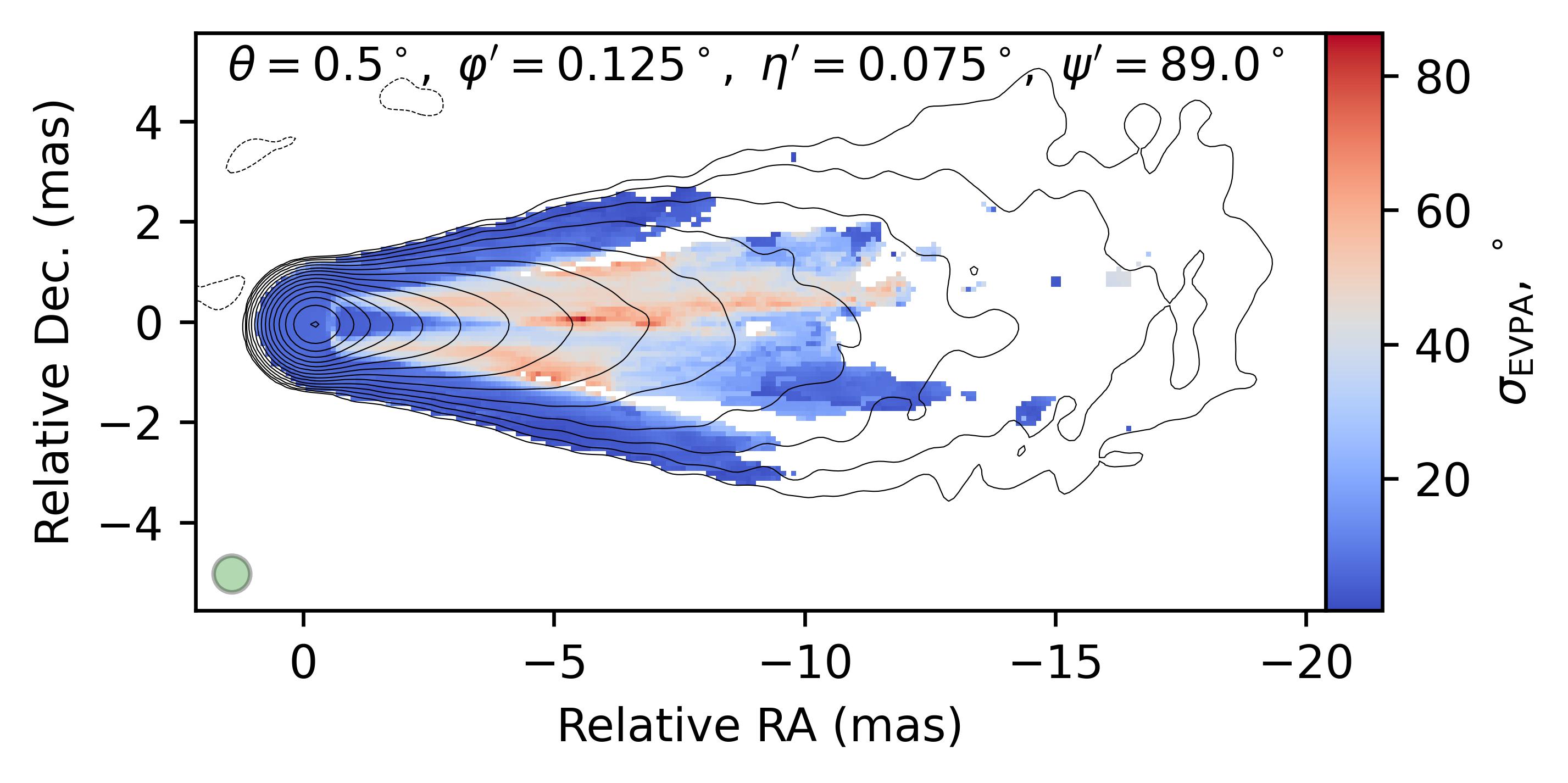}
\caption{Synthetic single-epoch (left) and stacked (right) VLBI maps of linear polarization of the jet. The notations are the same as in Fig.~\ref{fig:sim_vlbi_jet}. The model parameters are as follows: a conical jet with a true half-opening angle of $\phi'=0.125^{\circ}$, whose precession axis is inclined to the line of sight at $\theta=0.5^{\circ}$, precessing with an amplitude of $\eta'=0.075^{\circ}$; the pitch angle of the helical magnetic field is $\psi'=89^{\circ}$.
\label{fig:sim_pitch89_los05}}
\end{figure}

\begin{figure}
\includegraphics[width=0.47\columnwidth]{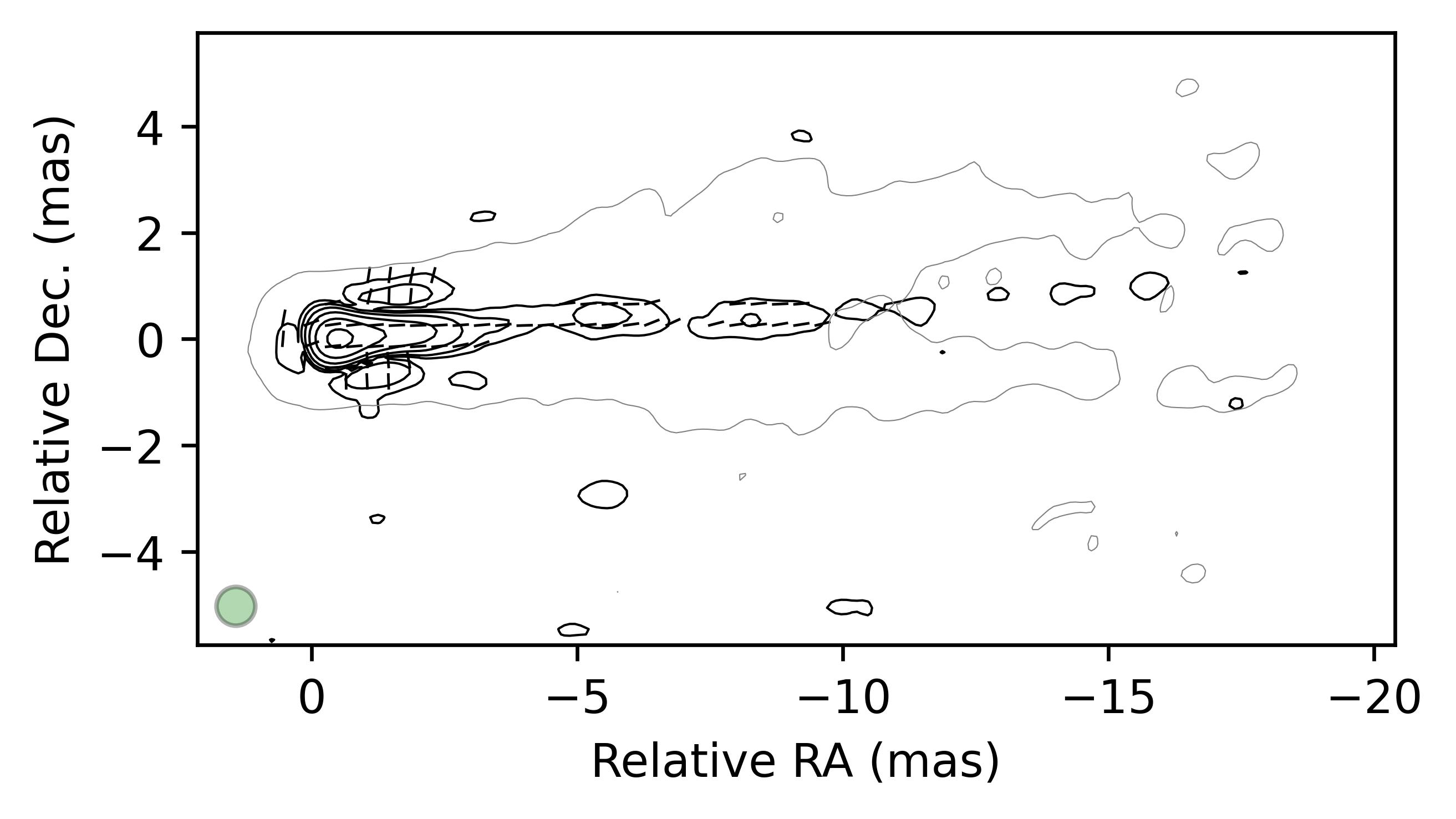}
\includegraphics[width=0.49\columnwidth]{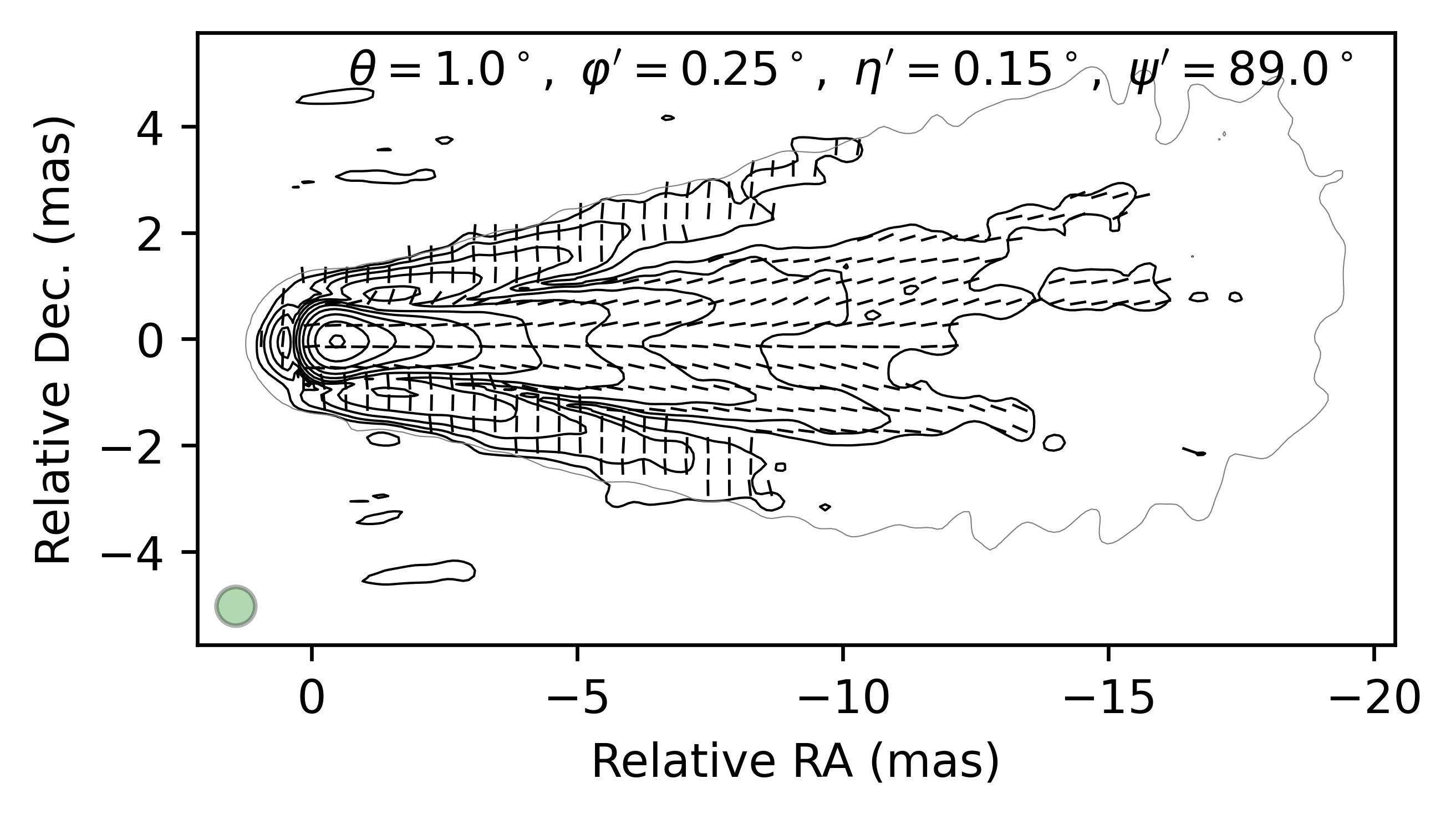}
\includegraphics[width=0.47\columnwidth]{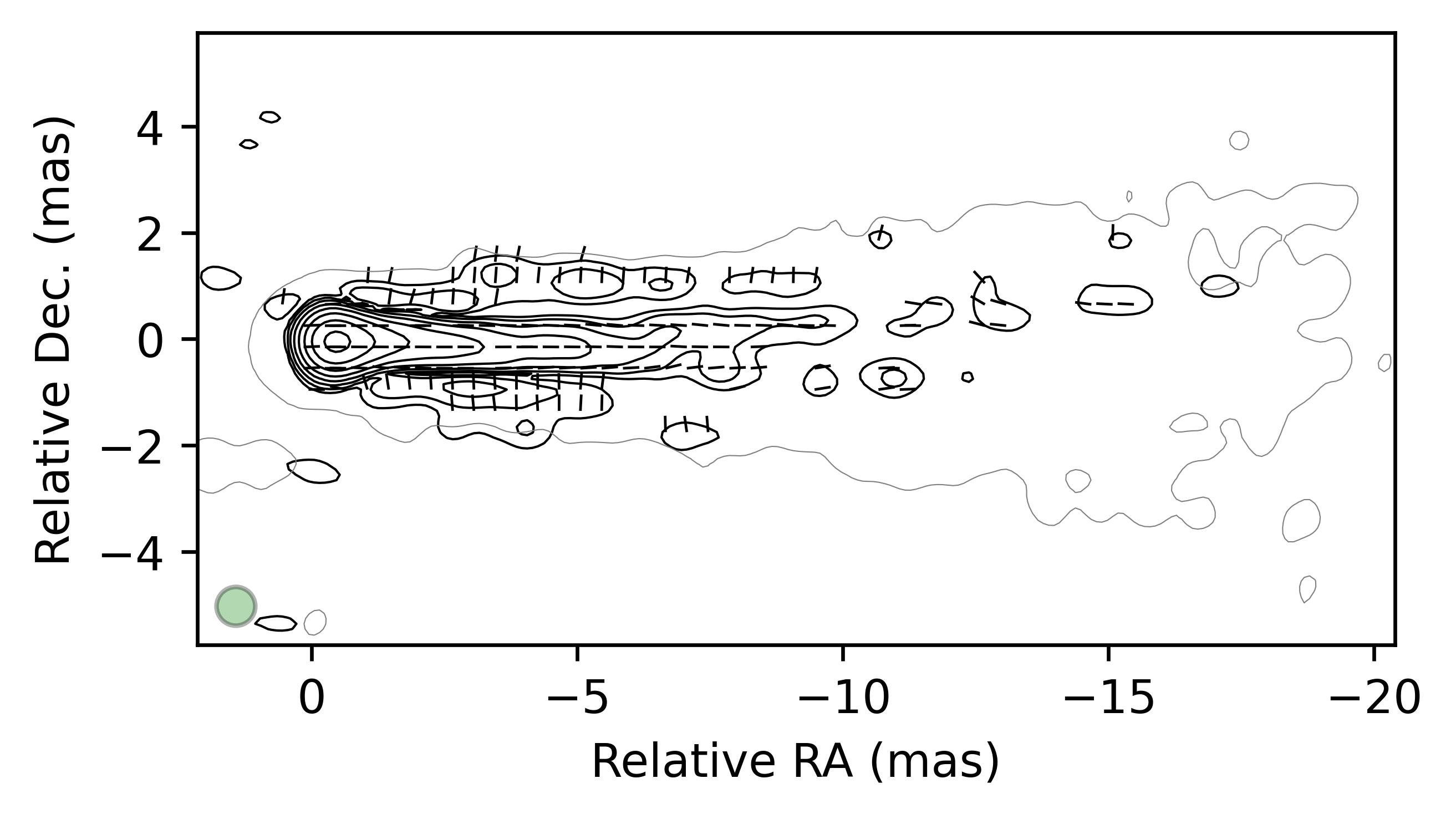}
\includegraphics[width=0.50\columnwidth]{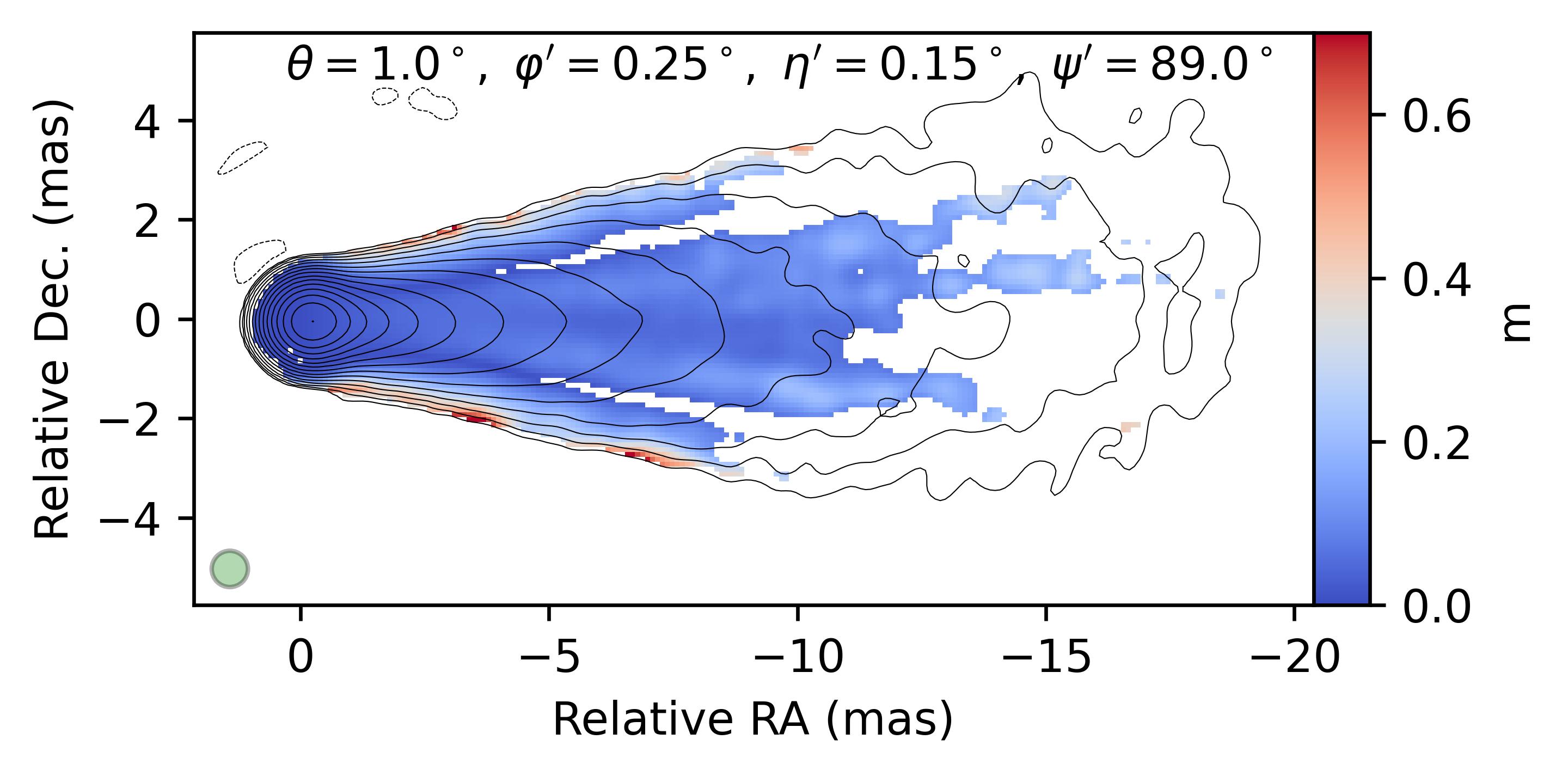}
\includegraphics[width=0.47\columnwidth]{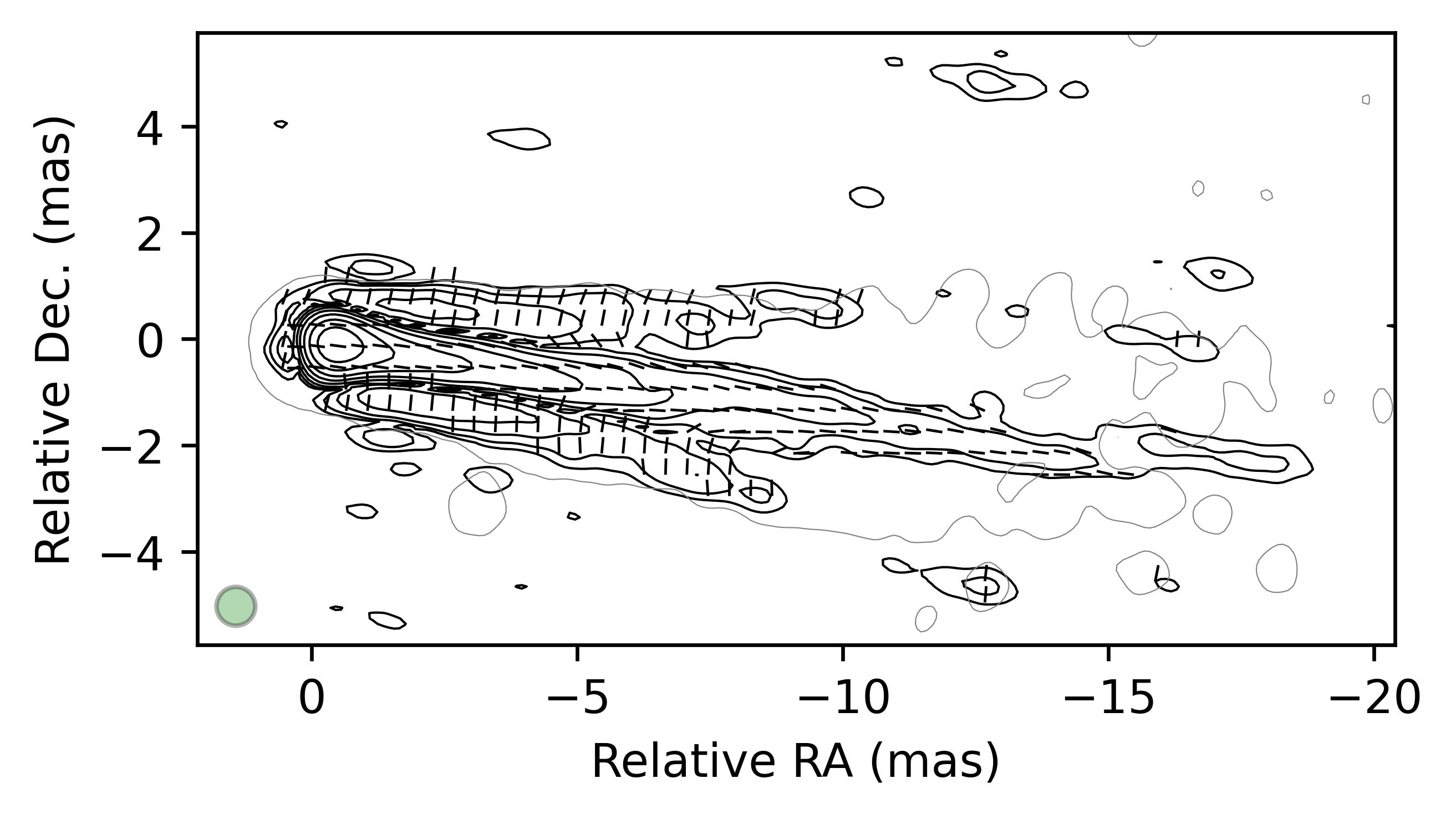}
\includegraphics[width=0.51\columnwidth]{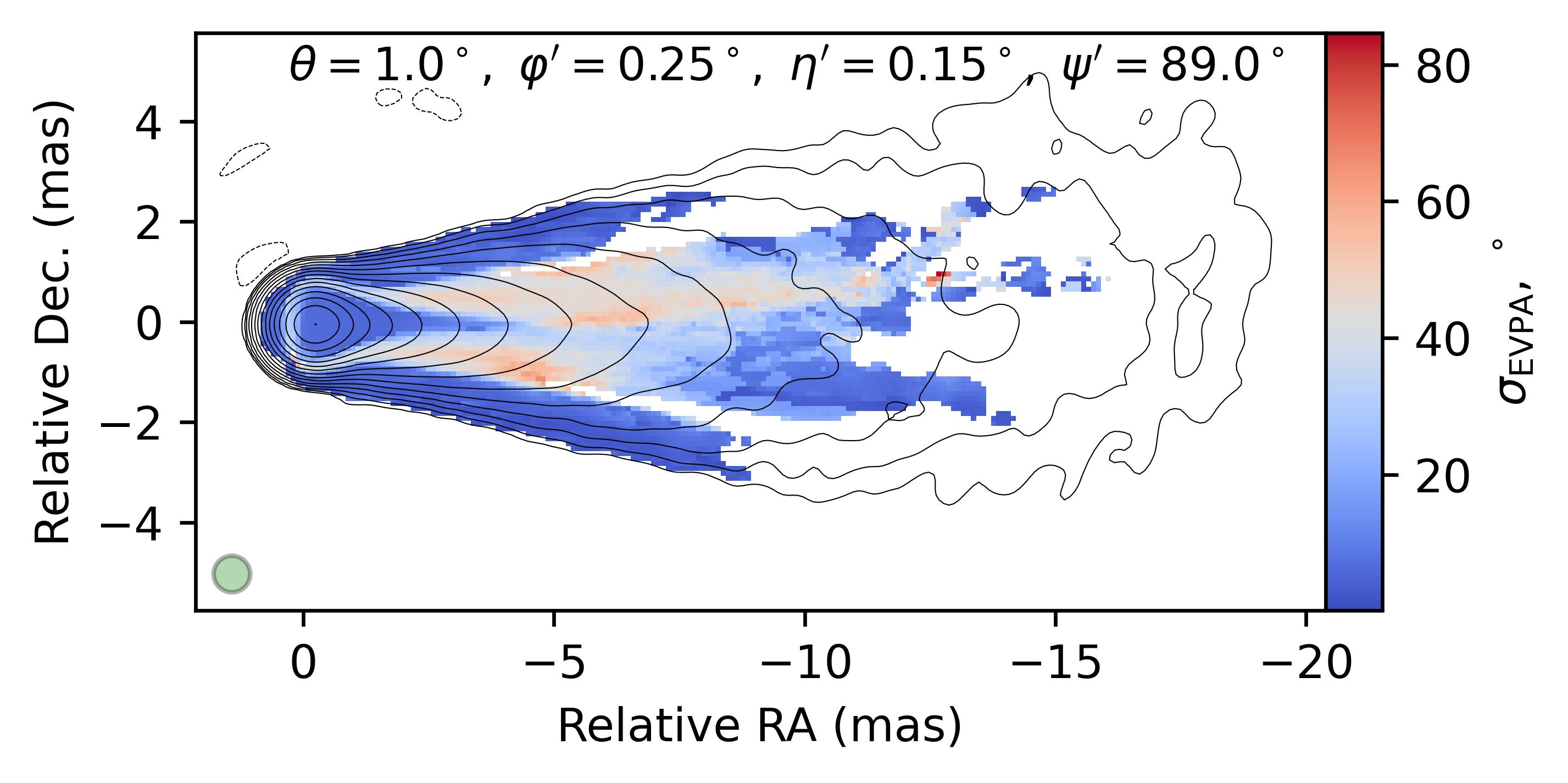}
\caption{Synthetic single-epoch (left) and stacked (right) VLBI maps of linear polarization of the jet. The notations are the same as in Fig.~\ref{fig:sim_vlbi_jet}. The model parameters: a conical jet with a true half-opening angle of $\phi'=0.25^{\circ}$, whose precession axis is inclined to the line of sight at $\theta=1^{\circ}$, precessing with an amplitude of $\eta'=0.15^{\circ}$; the pitch angle of the helical magnetic field is $\psi'=89^{\circ}$.
\label{fig:sim_pitch89_los1}}
\end{figure}

The produced stacked maps are shown in Figure~\ref{fig:sim_vlbi_jet}, Figure~\ref{fig:sim_pitch89_los05}, and Figure~\ref{fig:sim_pitch89_los1}.
The polarized emission of the jet is characterized by a triple-peak structure: a spine with the EVPAs aligned with the outflow direction and two side regions with transverse direction of polarization. The EVPA distribution shows a ``fountain'' pattern. 
The lowest degree of polarization is observed between the central and side peaks, as well as along the jet spine. An increase of the polarization degree toward the jet edges is partially caused by systematic errors during the CLEAN procedure \citep{2023MNRAS.520.6053P}. 
It is obvious, that due to the bright spine, the central peak with longitudinal polarization in the stacked images is not efficiently smeared within the beam, even when using a sufficiently narrow apparent opening jet angle.

\begin{figure}
\includegraphics[width=0.5\columnwidth]{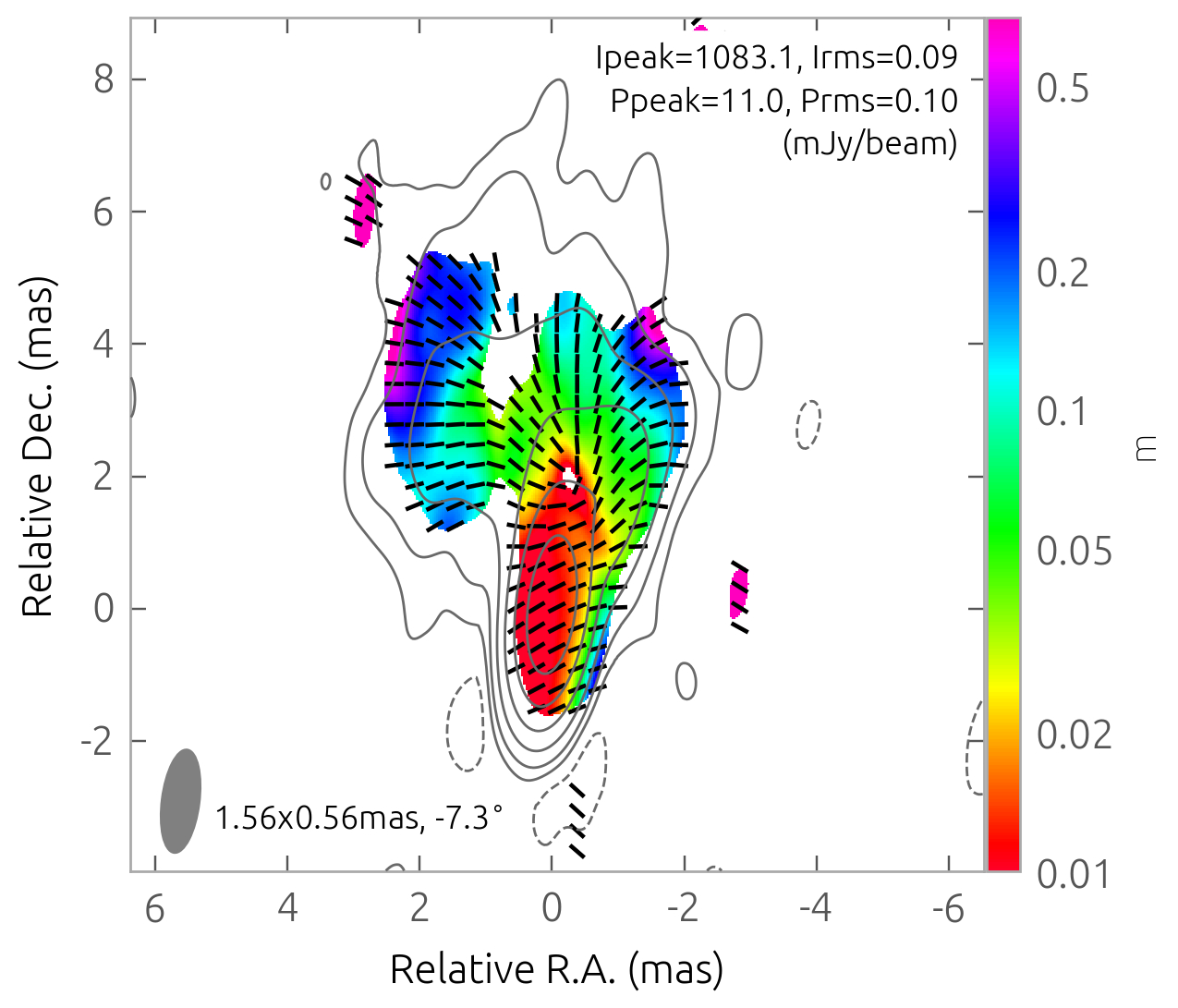}
\includegraphics[width=0.5\columnwidth]{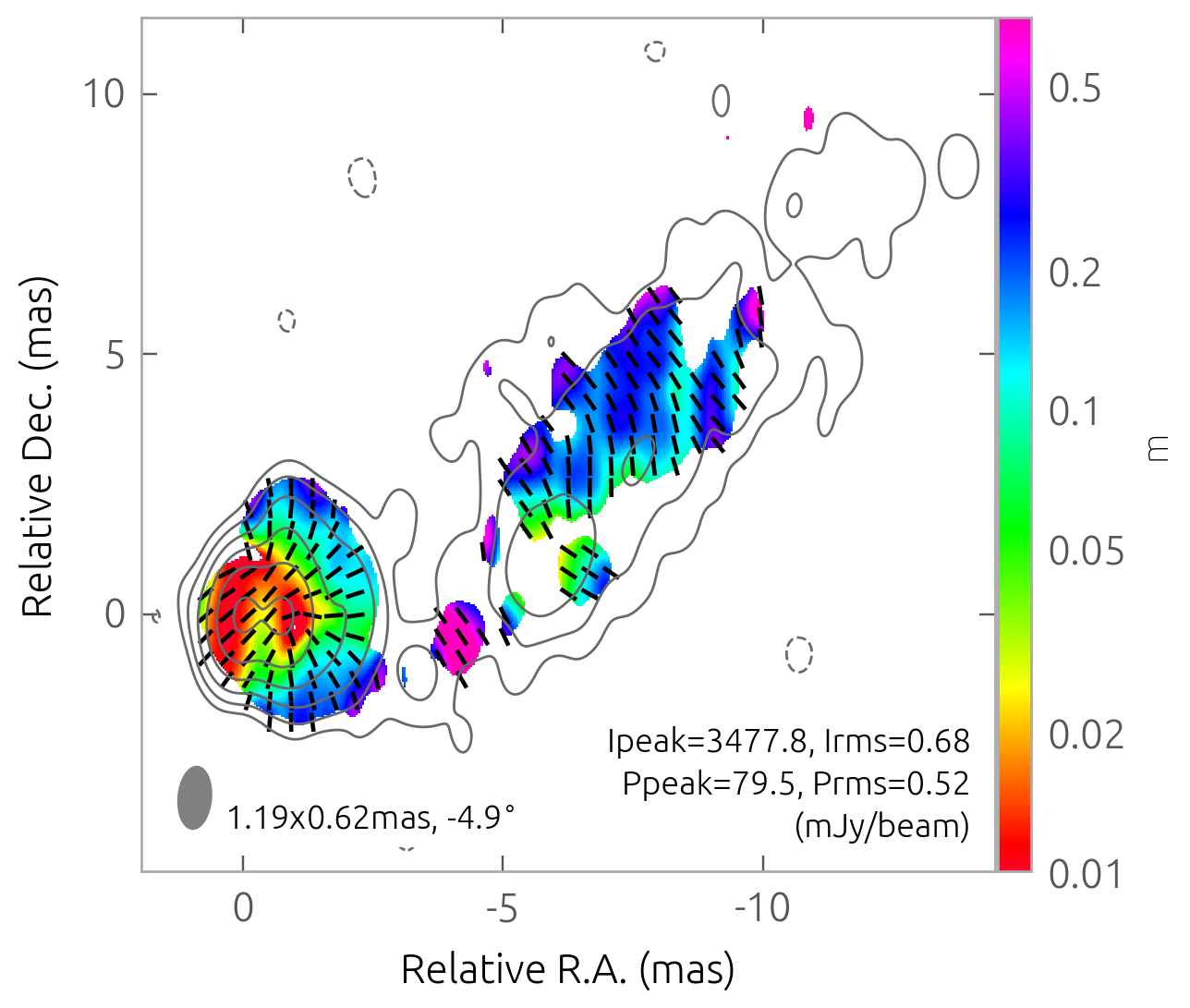}
\caption{Examples of observed single-epoch U-shaped polarization profiles. Maps of linear polarization of quasar 1920$-$211 on August 3, 2012 (left) and quasar 2251$+$158 on March 30, 2003 (right) at 15\,GHz. The color represents the degree of polarization (logarithmic scale), the contours show the total intensity, and the dashed lines depict the EVPAs uncorrected for the Faraday rotation. The data is taken from the website of the MOJAVE project:
\url{https://www.cv.nrao.edu/MOJAVE/}.
\label{fig:moj_1ep_uprofile}}
\end{figure}

In some sources, the U-shaped profile and the ``fountain'' EVPA distribution are observed in individual epochs, as for example shown in Fig.~\ref{fig:moj_1ep_uprofile}. Despite the fact that the model predicts the three-peak structure, effective averaging within the beam and the influence of noise can suppress the central peak with longitudinal polarization when assuming sufficiently narrow apparent jet opening angle. In the precession model, these epochs primarily correspond to phases with the largest inclination angle of the jet axis to the line of sight. The obtained result is robust with respect to the choice of the period and initial phase of precession.

\section{Discussion}

Assumption of a helical magnetic field in AGN jets has been widely and successfully used over the past twenty years to interpret observed polarization effects \citep{2001ApJ...561L.161G, 2002PASJ...54L..39A, zamaninasab_etal13, 2015MNRAS.450.2441G, 2016ApJ...817...96G}. 
This motivated as to chose a helical magnetic field model to simulate the linear polarization of the jet.
Due to the relatively bright spine with longitudinal polarization present in the considering models, the emission in the center of the precessing jet in the produced synthetic stacked maps is characterized by a non-zero degree of polarization. In other words, polarized radiation is not suppressed enough to form the characteristic observed polarization structure. Meanwhile, the reproduced structure and high variability of EVPAs in the center of the jet well consistent with observations.

Recently, \citet{2023MNRAS.523..887F} studied the intensity profiles of the AGN jets based on magnetohydrodynamic (MHD) semi-analytic simulations using models with constant angular velocity and a total electric current closed inside the jet, as well as different mechanisms of "heating" the emitting particles. The authors showed that, in assumption that all particles in the jet are emitting, the transverse intensity profile is dominated by a dense core.
Comparison with observations of a resolved jets shows the radiation from the core are to be suppressed. 
\citet{2023MNRAS.523..887F} demonstrate that, for example, this can be achieved through transverse stratification of the distribution of emitting particles, where only a small fraction of particles radiate up to a certain radius. One possible mechanism considered is the heating of emitting particles due to an Ohmic dissipation \citep{2005MNRAS.360..869L}.
The suppression of the core emission can also be related to the anisotropic distribution of accelerated particles observed in \cite{inv}. The conservation of the adiabatic invariant leads to the particles in the central regions of the outflow having a preferential motion direction along the magnetic field, dominated by the longitudinal component in this region.
\citet{2005MNRAS.360..869L} managed to explain the observed bimodality of the EVPA direction in jets in the hollow jet model, where the core is naturally suppressed. However, \citet{2023MNRAS.523..887F} show that although this model with a number density of emitting particles being proportional to the density of magnetic energy in the plasma structure is promising, it is unlikely as it leads to a strong dominance of the central core.

According to a recent structural analysis of 447 brightest radio-loud sources in the northern sky at a frequency of 15\,GHz \citep{2021ApJ...923...30L}, variations in the jet position angle is a common phenomenon, with an average amplitude of $10^{\circ}-50^{\circ}$ over a timescale of about ten years. For some sources, the variations reach $200^{\circ}$.
There are several scenarios that can explain such behavior. For example, orbital motion in a binary black hole system \citep{1980Natur.287..307B}.
A more widely discussed scenario is the Lense-Thirring effect \citep{1918PhyZ...19..156L, 1918PhyZ...19...33T} in a system with a rotating black hole. In this case, oscillations in the internal orientation of the jet are caused by the precession of the accretion disk, whose rotation axis is inclined to the black hole spin \citep{2004ApJ...616L..99C}.
For example, a study based on observations spanning 22 years and covering 170 individual epochs toward the radio galaxy M87 jet showed periodic changes in the jet position angle with an amplitude of $\approx10^{\circ}$ and a period of $T\approx11$ years on scales of  $\thicksim$600-2500 $r_{\rm g}$ \citep{cui2023}. Here, $r_{\rm g}$ is the gravitational radius defined as $r_{\rm g} = GM/c^2$, where $M$ is the mass of the black hole.
The periodicity of the optical light curve and variations in the jet position angle in quasar 3C~120 is sonsistent with the existence of jet precession caused by the Lense-Thirring effect, with a period of 12.3 years \citep{2004MNRAS.349.1218C}.
The same is shown for the jet in M81$^{*}$, the precession period of which is estimated at $\approx 7$ years \citep{2023arXiv230300603V}.
A detailed analysis of the orientation fluctuations of other jets also indicates their periodicity and possible connection with the precession. For example OJ~287\citep[$T\thicksim12$ years][]{2018MNRAS.478.3199B}, 3C~273 \citep[$T\thicksim16$years][]{1999A&A...344...61A}, 3C~279 \citep[$T\thicksim22$ year][]{1998ApJ...496..172A}, PG~1553+113 \citep{2020A&A...634A..87L}, 4C~38.41 \citep[$T=23 \pm 5$ years][]{2019ApJ...886...85A}, PKS~2131-021 \citep[$T\thicksim22$ years][]{2022ApJ...926L..35O}.

The existence of a jet rotation or its components can be traced through other observational characteristics. For instance, by the smooth rotation of the electric vector of linearly polarized optical emission up to 720$^{\circ}$ \citep{2010ApJ...710L.126M, 2008Natur.452..966M}. Alternatively, it can be observed through the modulation of the brightness temperature of jet components, depending on their distance from the core (Kravchenko et al. 2023, subm.). Small deviations in the direction of the jet in the source's central region and a small angle between the jet axis and the line of sight lead to significant variations in the Doppler factor and, consequently, in the brightness temperature \citep{2010ApJ...712L.160R}. This phenomenon is observed, for example, in quasars 3C~273 or 3C~279.
Another widely discussed model capable of causing oscillations in the jet position angle is the development of Kelvin-Helmholtz instability due to velocity shear at the jet edges \citep{2007ApJ...664...26H}. Detailed analyses capable to explain the observed jet structure have been presented, for instance, toward M87 \citep{2023arXiv230711660N}, 0836+710 \citep{2012ApJ...749...55P}, and 3C~273 \citep{2001Sci...294..128L}.

It is worth noting that in the used here model, some polarization patterns (e.g., with transverse EVPA in the VLBI core) arise at sufficiently small jet viewing angles $\theta_{\rm LOS} << 1/\Gamma$. Statistical analysis of flux-limited samples, including VLBI-kinematics, predicts a relatively small number of such objects \citep{1994ApJ...430..467V,2007ApJ...658..232C,2019ApJ...874...43L}. However, this result is based on a jet model with constant $\Gamma$, which does not account for transverse velocity stratification \citep{Kom09, BCKN-17,2019MNRAS.490.2200C}, and apparently, the necessary transverse stratification of the distribution of emitting particles \citep{2023MNRAS.523..887F}.

Another important factor to consider when examining the polarization structure is the presence of moving and stationary components in the jet. These can be either moving/standing shock waves compressing the magnetic field \citep{1980MNRAS.193..439L,1985ApJ...298..301H} and accelerating particles at the front, or plasma knots with an increased density of emitting particles \citep[so-called "plasmons," e.g.,][]{2016MNRAS.462.2747K, 2017MNRAS.468.4478L,2019MNRAS.485.1822P}. \cite{2023MNRAS.520.6053P} estimate the typical ratio of the distance traveled by a component along the jet during the observation time to the typical distance between components to be 1.5. In other words, components can significantly influence the polarization distribution patterns obtained from stacked maps. These issues will be investigated separately in our subsequent papers.

\section{Conclusions}

Recently, in the largest and most complete study of the linear polarization structure of Active Galactic Nuclei jets have been identified several types of polarimetric structures in the distribution of linear polarization and its variability \citep{2023MNRAS.520.6053P, 2023MNRAS.523.3615Z}. One of the most intriguing findings is the polarization distribution with the brightening toward the jet edges and an EVPA resembling a ``fountain'' pattern, exhibiting its high variability along the jet spine. Simple models of the magnetic field configuration are unable to explain this structure, while more complex models cannot explain high temporal variability in the jet center.

In this study, we propose a model to explain such pattern by considering a precessing jet observed at a small viewing angle, with a strongly twisted helical or toroidal magnetic field. 
Large variations in the jet position angle within $10^{\circ}-50^{\circ}$ detected by multi-epoch long-term monitoring programs in many AGN jets over decades serves as a the observational basis for the proposed model. This data can be successfully described by models of periodic oscillations with periods of about 10-20 years. Such behavior can be explained, for instance, with the Lense-Thirring effect causing precession in a system with a rotating black hole and inclined accretion disc.

We simulated multi-epoch polarimetric VLBI-observations of precessing jets using parameters derived from real observations. The produced synthetic stacked polarization VLBI maps exhibit a three-peak transverse profile. The maps of the EVPA distribution resemble a pattern reminiscent of a fountain, and the maps of its variability are characterized by high values along the jet spine.
The precession of the jet results in depolarization at the central jet regions due to the superposition of regions with different orientations of the electric vector projected onto the image plane. However, the presence of a bright core causes insufficient depolarization at the center of the jet, which is inconsistent with observations where the emission in the central peak is depolarized. Therefore, we conclude that it is necessary to consider models with the suppressed emission from the central core, consistent with the recent results from analytical MHD calculations \citep{2023MNRAS.523..887F}.

\section{Acknowledgements}
The authors thank Yu. A. Kovalev for valuable suggestions that
significantly improved the paper.
This study was supported by the Russian Science Foundation grant 20-72-10078, \url{https://rscf.ru/project/20-72-10078/}.

\vspace{20pt}

\begin{multicols}{2}
\bibliographystyle{mnras}
\small
\bibliography{litr}
\end{multicols}

\end{document}